\documentclass[11pt]{article}
\usepackage{jheppub}
\usepackage{amsmath}
\usepackage{amssymb}
\usepackage{braket}
\usepackage{bbold}
\usepackage{pifont}
\newcommand{\be}{\begin{eqnarray}}
\newcommand{\ee}{\end{eqnarray}}
\newcommand{\beqn}{\begin{eqnarray}}
\newcommand{\eeqn}{\end{eqnarray}}
\newcommand{\bsubeq}{\begin{subequations}}
\newcommand{\esubeq}{\end{subequations}}
\newcommand{\dd}{\mathrm{d}}
\newcommand{\nn}{\nonumber}
\newcommand{\Tr}{\mathrm{Tr}}

\newcommand{\gmn}{g_{\mu\nu}}
\newcommand{\fmn}{f_{\mu\nu}}
\newcommand{\bgmn}{\bar{g}_{\mu\nu}}
\newcommand{\bfmn}{\bar{f}_{\mu\nu}}
\newcommand{\emn}{\eta_{\mu\nu}}

\newcommand{\ha}{\hat\alpha}

\newcommand{\pmn}{P_{\mu\nu}}
\newcommand{\tpmn}{\tilde P_{\mu\nu}}
\newcommand{\gmnp}{g'_{\mu\nu}}
\newcommand{\fmnp}{f'_{\mu\nu}}
\newcommand{\ggmn}{G_{\mu\nu}}
\newcommand{\mmn}{M_{\mu\nu}}
\def\ph{\phantom}

\title{Extended Weyl Invariance in a Bimetric Model and Partial Masslessness}

\author[1]{S.F.~Hassan,}
\author[2]{Angnis~Schmidt-May,}
\author[3]{Mikael~von~Strauss}
\affiliation[1]{Department of Physics \& 
        The Oskar Klein Centre,\\
        Stockholm University, AlbaNova University Centre, 
        SE-106 91 Stockholm, Sweden}
\affiliation[2]{Institut f\"ur Theoretische Physik, Eidgen\"ossische
  Technische Hochschule Z\"urich\\ 
Wolfgang-Pauli-Strasse 27, 8093 Z\"urich, Switzerland}
\affiliation[3]{UPMC-CNRS, UMR7095,
 Institut d'Astrophysique de Paris, GReCO,\\
 98bis boulevard Arago, F-75014 Paris, France.}
 
\emailAdd{fawad@fysik.su.se}
\emailAdd{angniss@itp.phys.ethz.ch}
\emailAdd{strauss@iap.fr}

\abstract{We revisit a particular ghost-free bimetric model which is related to both partial masslessness (PM) and conformal gravity. Linearly, the model propagates six instead of seven degrees of freedom not only around de Sitter but also around flat spacetime. Nonlinearly, the equations of motion can be recast in the form of expansions in powers of curvatures, and exhibit a remarkable amount of structure. In this form, the equations are shown to be invariant under scalar gauge transformations, at least up to six orders in derivatives, the lowest order term being a Weyl scaling of the metrics. The terms at two-derivative order reproduce the usual PM gauge transformations on de Sitter backgrounds. At the four-derivative order, a potential obstruction that could destroy the symmetry is shown to vanish. This in turn guarantees the gauge invariance to at least six-orders in derivatives. This is equivalent to adding up to 10-derivative corrections to conformal gravity. More generally, we outline a procedure for constructing the gauge transformations order by order as an expansion in derivatives and comment on the validity and limitations of the procedure. We also discuss recent arguments against the existence of a PM gauge symmetry in bimetric theory and show that, at least in their present form, they are evaded by the model considered here. Finally, we argue that a bimetric approach to PM theory is more promising than one based on the existence of a fundamental PM field.}


\begin{document} 
\maketitle
\flushbottom

\section{Motivation and summary of results}

The two major unresolved issues in General Relativity (GR) are the
non-renormalisability of the theory and the cosmological constant
problem. One expects that a theory of spin-2 fields with more symmetry
than GR will be better behaved in these respects. One such theory is
conformal gravity \cite{Bach} defined by the action,
\be
S_{\mathrm{CG}}=\int\dd^4x\sqrt{g}\,W_{\mu\nu\rho\sigma}W^{\mu\nu\rho\sigma}\sim
\int\dd^4x\sqrt{g}\,\left(R_{\mu\nu}R^{\mu\nu}-\tfrac{1}{3}R^2\right)\,
\ee
in terms of the Weyl (W) or Ricci (R) tensors; the two formulations differ only by the topological Euler density. 
In addition to diffeomorphism invariance this action is invariant 
under a Weyl scaling of the metric. This is a $4$-derivative theory and
propagates 6 modes, consisting of a massless spin-2 field and 4 ghost
modes \cite{Stelle:1976gc,Kaku:1977pa,Ferrara:1977mv,Fradkin:1981iu, 
Riegert:1984hf,Maldacena:2011mk}. Despite having an extra symmetry as compared to GR, the presence of the ghost instability makes the
theory less attractive.\footnote{Nonlinear infinite derivative generalisations geared towards avoiding the ghost problem have been considered in e.g.~\cite{Tseytlin:1995uq, Biswas:2005qr,Biswas:2011ar}, but the Weyl symmetry is then generically destroyed.}   

Another theory with a novel gauge symmetry is the linear partially
massless (PM) theory of a massive spin-2 field in an Einstein-de
Sitter (EdS) background
\cite{Deser:1983mm,Deser:2001pe,Deser:2001us}. Due to the symmetry,
the spin-2 field propagates 4 (instead of the usual 5)
polarisations. To this, one can add a linear massless graviton
(regarding the PM field itself as the graviton requires giving up
general covariance). The spectrum is now similar to that of conformal
gravity linearised around EdS backgrounds
\cite{Maldacena:2011mk}. However, while ghost-free, this theory is
non-interacting and exists only around special (EdS) backgrounds. The
question is if there exists a unitary theory of interacting spin-2 
fields with a PM-like gauge symmetry.\footnote{To be of relevance to
  the renormalisation and the cosmological constant problems, the PM
  symmetry must also affect the physical gravitational field.} The
first investigations involved explicit construction of cubic
interaction vertices
\cite{Zinoviev:2006im,Zinoviev:2013hac,Joung:2012rv}, but problems are
encountered at the quartic level.

On the other hand, a nonlinear setup with the right general features
is the ghost-free bimetric theory \cite{Hassan:2011zd}.  There exists
a unique bimetric model that, when linearised around EdS backgrounds,
contains a massless and a partially massless spin-2 field, exhibits
linear PM gauge symmetry and also up to
realises the {\it global} part of the PM transformations
nonlinearly~ \cite{Hassan:2012gz, Hassan:2012rq}. But this is far from establishing the full gauge
invariance of the nonlinear model. It turns out that the model also
exhibits features of conformal gravity. In a derivative expansion, it
reproduces the conformal gravity equation of motion at the lowest
order \cite{Hassan:2013pca} and, hence, is invariant under local Weyl 
scalings to this order. In this paper we explore this property
beyond the lowest order and discuss its relevance to the presence or
absence of PM symmetry. We show that the Weyl scalings can be extended
at least up to six-derivative terms to maintain the symmetry of the
equations. When restricted to the linear theory in EdS backgrounds,
one recovers the standard PM transformations. The bimetric model
provides a unified description of linear PM theory and conformal
gravity up to higher derivative corrections. More work is needed to
find if this structure can be extended to all orders.

Another potential setup for PM symmetry is nonlinear massive gravity
\cite{deRham:2010kj}, which is ghost-free \cite{Hassan:2011hr} and can
also be formulated around any background, including EdS spacetimes 
\cite{Hassan:2011vm,Hassan:2011tf,Hassan:2012qv}. A specific massive
gravity model, with a fixed dS reference metric, was identified in
\cite{deRham:2012kf} as exhibiting PM symmetry in the decoupling
limit. However, it was soon argued that this action lacks the gauge
symmetry at the nonlinear level \cite{Deser:2013uy,deRham:2013wv,
  Deser:2013gpa}.\footnote{The analysis of constraints in
  \cite{Comelli:2014xga} rules out massive gravity with 4
  polarisations for a Minkowski reference metric which is
  consistent with the linear PM theory.}

There have also been arguments against PM symmetry in the nonlinear
bimetric model. Since massive gravity models can be regarded as a
limit of bimetric models around given solutions, it has been argued,
for instance in \cite{Deser:2013uy,Deser:2013gpa}, that ruling out PM
symmetry in massive gravity, implies the same for the bimetric
model.\footnote{The massive gravity limit of the bimetric model in
  \cite{Hassan:2012gz} does not seem to reproduce the massive gravity
  PM candidate identified in \cite{deRham:2012kf}, but is consistent
  with the parameters values found in the subsequent work
  \cite{deRham:2013wv}.} However, \cite{Hassan:2014vja} argued that if
the bimetric model indeed had a PM symmetry, it would most likely be
destroyed in the massive gravity limit. These results are supported by
the findings in the present work. Other arguments are discussed in the
last section.  Recent work on the subject of partial masslessness
includes \cite{Joung:2014aba,Alexandrov:2014oda,Hinterbichler:2014xga,
  Garcia-Saenz:2014cwa}.

\paragraph{Summary of results:} This paper does not directly deal with
PM symmetry and the possibility of its realisation in a nonlinear
theory. Rather, we explicitly investigate a particular bimetric model
in a derivative expansion. On eliminating one of the metrics between
the two equations, one obtains the 4-derivative conformal gravity
equation of motion which is invariant under Weyl scaling of the
metrics \cite{Hassan:2013pca}. Here we investigate the possibility of
extending the zero-derivative gauge symmetry to higher orders in
derivatives. The structure of the equations enables us to find a
prescription for constructing the gauge transformations order by order,
\be
\Delta\gmn&=&\Delta_{(0)}\gmn+\Delta_{(2)}\gmn+\Delta_{(4)}\gmn 
+\Delta_{(6)}\gmn+\hdots\,,\nn\\ \Delta\fmn&=&\Delta_{(0)}\fmn 
+\Delta_{(2)}\fmn+\Delta_{(4)}\fmn+\Delta_{(6)}\fmn+\hdots\,, 
\label{GTsum}
\ee 
which leave the equations of motion invariant on-shell up to sixth
order in derivatives. Here, $\Delta_{(0)}\gmn=\phi\gmn$ and 
$\Delta_{(0)}\fmn=\phi\fmn$ are Weyl scalings of the metrics and each
$\Delta_{(2n)}$ contains terms with $2n$ derivatives. The prescription
insures that terms with $2n=(4m+2)$ derivatives (for integer $m$) in
the transformation can always be constructed. But the existence of the
$2n=4m$-derivative terms requires the model to 
satisfy certain conditions. It is explicitly demonstrated that these
conditions are satisfied for the four-derivative term and thus we are
able to show the invariance of the equations of motion up to six
orders in derivatives. This is equivalent to considering up to
$10$-derivative corrections to the $4$-derivative Bach equation.

On Einstein-de Sitter backgrounds, the $2$-derivative terms in the
transformations of the nonlinear metrics reduce to the well-known
gauge transformations for a partially massless spin-2
perturbation. This suggests that \eqref{GTsum} may be considered as
potential extensions of the linear PM gauge symmetry. But wether such
a gauge symmetry really exists, remains to be seen. In this setup,
the metrics are not restricted to de Sitter backgrounds and hence we
can also study the behaviour of the model around other
backgrounds. Unlike its massive gravity counterpart, the bimetric
model also has flat background solutions which enable us to
extend the notion of linear partial masslessness to Minkowski
space. One finds that around flat backgrounds no decomposition into
spin-2 mass eigenstates exists, instead, the spectrum coincides with
the conformal gravity spectrum around flat spacetime, {\it i.e.,} two
massless tensors and one massless vector. The results are discussed in
more detail in the text and in section \ref{sec:sum}.

The paper is organised as follows. In section~\ref{sec:rev} we review
details of ghost-free bimetric theory, the perturbative expansion of
its equations of motion, and the emergence of the Bach
equation in a particular bimetric model. Section~\ref{sec:rel}
considers this model at the quadratic level and the relation to both
linear PM theory and linear conformal gravity, including the flat
background case. In section~\ref{sec:sym} we study the perturbative
expansions of equations in this model and outline the procedure for
constructing the higher-derivative terms in the gauge transformations
of the metrics. We establish the invariance to sixth order in
derivatives, and discuss the relation to PM transformations, as well as
the limitations of the procedure. Our results are discussed in
section~\ref{sec:sum}, where we also comment on various counter
arguments and no-go results in the recent literature. Some technical
details are provided in the appendices.
 
\section{Review of the derivative expansion in bimetric theory}
\label{sec:rev}

Here we briefly review the structure of ghost-free bimetric theory and
outline how to derive the higher-curvature expansions of its equations
of motion. The main results are summarised in this sections; some more
details can be found in appendix~\ref{app:details}.

\subsection{The bimetric action}

The ghost-free action for two spin-2 fields $\gmn$ and $\fmn$ with  
non-derivative interactions is \cite{Hassan:2011zd}, 
\be
{\cal S}(g,f)=\int\dd^4 x~\left[m_g^2\sqrt{g}~R(g)+m_f^2\sqrt{f}~R(f)
-2m^4 \sqrt{g}~V(\sqrt{g^{-1}f}\,)\,\right],
\label{act}
\ee where, $m_g$ and $m_f$ are the two Planck masses and $m$ is an
additional mass scale. The potential is given in terms of a
square-root matrix $S\equiv\sqrt{g^{-1}f}$ as, 
\beqn
V(S)=\sum_{n=0}^{4}\tfrac{\beta_n}{n!(4-n)!}
\epsilon_{\mu_1\cdots\mu_n\lambda_{n+1}\cdots\lambda_{4}}
\epsilon^{\nu_1\cdots\nu_n\lambda_{n+1}\cdots\lambda_{4}}
S^{\mu_1}_{~~\nu_1}\cdots S^{\mu_n}_{~~\nu_n}\, 
\equiv \sum_{n=0}^{4}\beta_ne_n(S).  
\eeqn 
where $\beta_n$ are five interaction parameters. The $e_n(S)$ are the
elementary symmetric polynomials of the eigenvalues of $S$; for their
definitions see appendix~\ref{app:detbim}.\footnote{For a
  non-dynamical $f_{\mu\nu}=\eta_{\mu\nu}$ and a restricted set of
  $\beta_n$, $V(S)$ becomes the massive gravity potential first
  proposed in \cite{deRham:2010ik, deRham:2010kj} and proven to be
  free of the Boulware-Deser ghost at the nonlinear level in
  \cite{Hassan:2011hr,Hassan:2011tf, Hassan:2012qv}. } 

The fact that both $\gmn$ and $\fmn$ in \eqref{act} have
Einstein-Hilbert kinetic terms is related to the invariance of the  
potential under the interchanges $g\leftrightarrow f$ and 
$\beta_n\leftrightarrow \beta_{4-n}$. This invariance directly follows 
from the identity \cite{Hassan:2011zd},
\be
\sqrt{g}\sum_{n=0}^{4}\beta_n\,e_n(S)=
\sqrt{f}\sum_{n=0}^{4}\beta_{4-n}\,e_n(S^{-1})\,.
\ee
A Hamiltonian analysis shows that this theory propagates 7 modes
\cite{Hassan:2011zd,Hassan:2011ea} and no Boulware-Deser ghost
\cite{Boulware:1972zf,Boulware:1973my}. Around backgrounds of the type
$\bar f_{\mu\nu}=c^2\bar g_{\mu\nu}$, these modes combine into
massless and massive spin-2 fluctuations with, respectively, 2 and 5
polarisations \cite{Hassan:2012wr}.  In general, a bimetric theory can
be interpreted as describing a gravitational metric $g_{\mu\nu}$ (with
standard matter couplings) in the presence of an extra spin-2 field
$f_{\mu\nu}$. Then, for $m_f<<m_g=M_p$, the physical metric
$g_{\mu\nu}$ is a mostly massless field (in contrast to the massive
gravity limit $m_f\rightarrow\infty$) \cite{Hassan:2012wr,
  Akrami:2015qga}.

\subsection{Perturbative expansion of bimetric equations}\label{sec:PM}

The equations of motion obtained on varying the action \eqref{act}
with respect to $\gmn$ and $\fmn$ are of the form,
\begin{subequations}
\label{eom}
\begin{align}
g\mbox{-eom}:\qquad
&\tfrac{1}{\mu^2}\left(R^\mu_{~\nu}-\tfrac{1}{2}\delta^\mu_{~\nu}R\right) 
+ V^\mu_{~\nu}(S)=0\,,   \label{g-eom}\\
f\mbox{-eom}:\qquad
&\tfrac{\alpha^2}{\mu^2}\left(\tilde R^\mu_{~\nu}-\tfrac{1}{2}
\delta^\mu_{~\nu} \tilde R\right)+\tilde V^\mu_{~\nu}(S^{-1})=0\,. 
\label{f-eom}  
\end{align}
\end{subequations}
In these expressions, $R_{\mu\nu}$ and $\tilde R_{\mu\nu}$ are the
curvatures of $\gmn$ and $\fmn$, respectively, and $V_{\mu\nu}$ and
$\tilde V_{\mu\nu}$ are the corresponding interaction contributions,
explicitly given in appendix~\ref{app:detbim}. Moreover,
$S=\sqrt{g^{-1}f}$. In the $g$-equation, the first index is raised by
$g^{\mu\nu}$, and in the $f$-equation by $f^{\mu\nu}$. To simplify the
expressions, we use the notation,   
\be
\mu^2\equiv \frac{m^4}{ m_g^2} \,, \qquad
\alpha\equiv\frac{m_f}{m_g}\,, 
\label{mualpha}
\ee

The $g$-equation depends on $\fmn$ (through $S$) but does not contain 
derivatives of $\fmn$. Hence, in principle, it can be algebraically
solved for $\fmn$ in terms of $R_{\mu\nu}(g)/\mu^2$. For example, when
$\beta_0$ and $\beta_1$ are the only non-vanishing parameters, the
potential is $V=\beta_0+\beta_1 \Tr(S)$ and its variation gives
$V^\mu_{~\nu}(S)=\beta_0\delta^\mu_{~\nu}+\beta_1(S^\mu_{~\nu}-
\delta^\mu_{~\nu}\Tr(S))$. In this case, \eqref{g-eom} yields the
exact expression,
\begin{align}
S^\mu_{~\nu}=  -\tfrac{\beta_0}{3\beta_1}\delta^\mu_{~\nu}+
\tfrac{1}{\beta_1\mu^2}g^{\mu\rho}P_{\rho\nu}\,,
\end{align}
where $P_{\mu\nu}$ is the (scaled) Schouten tensor of $\gmn$,
\be
P_{\mu\nu}=R_{\mu\nu}-\tfrac{1}{6}\,\gmn R\,.
\ee
Since $\fmn=g_{\mu\rho}(S^2)^\rho_{~\nu}$, one immediately obtains, 
\be
\fmn= \tfrac{\beta_0^2}{9\beta_1^2}\gmn-
\tfrac{2\beta_0}{3\beta_1^2\mu^2}P_{\mu\nu} +
\tfrac{1}{\beta_1^2\mu^4}P_{\mu\rho}g^{\rho\sigma}P_{\sigma\nu}\,.
\ee
This is simply a rewriting of the $g$-equation, but does not involve
solving it as a differential equation.  Using this in the $f$-equation
\eqref{f-eom}, gives a higher derivative equation for $\gmn$ alone.   

For generic $\beta_n$ parameters in \eqref{g-eom}, it is not easy to
obtain an exact solution for $S$, and hence for $f=gS^2$. But, for
small $R_{\mu\nu}/\mu^2$, it is always possible to find a perturbative
solution \cite{Hassan:2013pca}. At lowest order, {\it i.e.},
neglecting $R_{\mu\nu}/\mu^2$, this gives $\fmn=a^2\gmn+\cdots$, with
the constant $a$ determined by the polynomial equation
$\left.V^\mu_{~\nu}\right|_{S=a\mathbb{1}}=0$. Curvature corrections
to this can be systematically computed and one arrives at an
expression for $\fmn$ of the form \cite{Hassan:2013pca},
\be
\label{fsol}
\fmn=a^2\gmn+\frac{b}{\mu^2}P_{\mu\nu}+\frac{c_1}{\mu^4}P^2_{\mu\nu}
+\frac{c_2}{\mu^4}\left[ \tfrac{1}{3}e_2(P)\gmn-PP_{\mu\nu}\right]+
\mathcal{O}\left(\tfrac{P^3}{\mu^6}\right)\,, 
\ee 
where indices on the right-hand side are contracted with $\gmn$.  The
coefficients $a$, $b$ and $c_n$ are given in terms of bimetric
parameters in appendix~\ref{app:obtHC}. Equation~\eqref{fsol} is a
rewriting of the $g$-equation \eqref{g-eom} and is satisfied by the
compatible solutions of the latter perturbatively,\footnote{Let us
  comment on the generality of such expansions. If $\gmn$ is an
  Einstein metric, then $P_{\mu\nu}= (\Lambda/3) \gmn$ and
  \eqref{fsol} implies $\fmn=c^2\gmn$, for some constant $c^2$. On the
  other hand, in generic bimetric models when one metric is Einstein,
  the other one is Einstein too, but the two metrics are not
  necessarily proportional to each other. Obviously, such
  non-proportional Einstein solutions are not captured by
  \eqref{fsol}. However, in classes of bimetric models, the
  $\beta_i$-models (where only one $\beta_i$ out of $\beta_1, \beta_2,
  \beta_3$ is non-zero), when either metric is Einstein, one
  necessarily has $\fmn=c^2\gmn$, as implied by \eqref{fsol}
  \cite{Hassan:2014vja}. In this paper we will work with a
  $\beta_2$-model. \label{footnote:beta_i}} as long as curvatures are
small compared to the mass scale $\mu=m^2/m_g$. In principle, the
coefficients in~\eqref{fsol} can be determined to arbitrary order, and
terms with $2n$ derivatives are suppressed by $\mu^{2n}$.

Alternatively, we can obtain a solution for $\gmn$ from the
$f$-equation \eqref{f-eom}. It has a very similar form,      
\be
\label{gsol}
\gmn=\tilde{a}^2\fmn+\frac{\tilde{b}}{\tilde{\mu}^2}
\tilde{P}_{\mu\nu}+\frac{\tilde{c}_1}{\tilde{\mu}^4}\tilde{P}^2_{\mu\nu}
+\frac{\tilde{c}_2}{\tilde{\mu}^4}\left[\tfrac{1}{3}e_2(\tilde{P})\fmn
-\tilde{P}\tilde{P}_{\mu\nu}\right]+\mathcal{O}\left(\tfrac{\tilde{P}^3}
{\tilde{\mu}^6}\right)\,,
\ee
where $\tilde{P}_{\mu\nu}\equiv\pmn(f)$ is the Schouten tensor for
$\fmn$. Indices are contracted with $\fmn$ and the suppressing mass
scale is now $\tilde{\mu}=\mu/\alpha$.

For generic models, the two expansions \eqref{fsol} and \eqref{gsol}
are not simultaneously valid. For example, \eqref{fsol} at the lowest
order reads $\fmn=a^2\gmn+\cdots$, whereas \eqref{gsol} is of the form
$\gmn= \tilde{a}^2\fmn+\cdots$. Obviously, a necessary condition for
the validity of both expansions is that $\tilde{a}^2=a^2$ (which can
be satisfied by fixing one of the $\beta_n$). The model we consider in
this paper satisfies this property. 

It is possible to use the expression \eqref{fsol} for $\fmn$ to
eliminate it from the $f$-equation (\ref{f-eom}). This yields the
following higher derivative equation for $\gmn$,   
\begin{align}\label{fEq2nd}
&x_{00}\gmn+\tfrac{x_{10}}{\mu^2}
\mathcal{G}_{\mu\nu}
+\tfrac{x_{11}}{\mu^2}\,P_{\mu\nu} 
+\tfrac{x_{20}}{\mu^4}B_{\mu\nu}\nn\\
&+\tfrac{x_{21}}{\mu^4}\Big[(s_1+2s_2)P_\mu^{~\rho}
P_{\rho\nu} -2s_2PP_{\mu\nu}-\tfrac{s_2}{3}\gmn\left(P_{\rho\sigma}
P^{\rho\sigma}-P^2\right)\Big]
\nn\\
&-\tfrac{x_{22}}{\mu^4}\left[3P P_{\mu\nu}
-2P_\mu^{~\rho} P_{\rho\nu} -\frac{1}{2}g_{\mu\nu}(P^2-P^{\alpha\beta}
P_{\alpha\beta})\right]+\mathcal{O}\left(\tfrac{P^3}{\mu^6}\right) 
=0\,.
\end{align}
The coefficients $x_{mn}$ and $s_n$ are given in appendix
\ref{app:obtHC}, and $\mathcal{G}_{\mu\nu}=R_{\mu\nu}-\tfrac1{2}\gmn
R$ is the Einstein tensor of $\gmn$. We have collected some of the
four-derivative terms into the Bach tensor \cite{Bach},  
\be\label{bacht}
B_{\mu\nu}=-\nabla^2P_{\mu\nu}&-\nabla_\mu\nabla_\nu P^\rho_{~\rho}
+\nabla_\rho\nabla_\mu P^\rho_{~\nu}+\nabla_\rho\nabla_\nu P^\rho_{~\mu}
-2P_\mu^{~\rho}P_{\rho\nu}+\tfrac1{2}\gmn P^{\rho\sigma}P_{\rho\sigma}\,.
\ee
Note that the Bach equation, $B_{\mu\nu}=0$, is the equation of motion
for conformal gravity.

Equation \eqref{fEq2nd} can be rewritten as an Einstein equation for
$\gmn$ with higher derivative corrections. The highest number of
derivatives on the same $\gmn$ is four; these appear in $B_{\mu\nu}$
and in some higher order terms which all arise from expanding the
Einstein tensor $\tilde{\cal G}_{\mu\nu}$ of $\fmn$. All other terms,
including the corrections, contain a maximum of two derivatives on
$\gmn$, but in higher powers. This is consistent with the fact that we
need to specify four initial conditions in the original bimetric
equations.

To summarise, we have re-expressed equations \eqref{g-eom} and
\eqref{f-eom} as \eqref{fsol} and \eqref{fEq2nd}. The perturbative
equivalence between these two sets of equations holds algebraically
and the solutions of the earlier set satisfy \eqref{fEq2nd}
perturbatively (subject to the comment in footnote
\ref{footnote:beta_i}). For the purpose of this paper, it is not
necessary that these two sets are also equivalent as differential
equations. In particular, studying the symmetry properties of the
equations involves only algebraic manipulations and, for such
purposes, the two sets of equations can be treated perturbatively
equivalent.

\subsection{A model with a possible gauge symmetry}

Let us consider a particular bimetric model that leads to an equation  
\eqref{fEq2nd} for $\gmn$ with, 
\be 
x_{00}=x_{10}=x_{11}=0\,, 
\ee 
so that the equation starts at fourth order in derivatives.  Using
the expressions for the $x_{mn}$, it can easily be shown that these
conditions uniquely fix the bimetric interaction parameters to the
following values,  
\be
\label{Pmpar} \beta_1=\beta_3=0\,,\qquad
\alpha^4\beta_0=3\alpha^2\beta_2=\beta_4\,.  
\ee 
It turns out that this choice of parameters also sets
$x_{21}=x_{22}=0$ and thus specifies a bimetric theory whose equations   
of motion imply \cite{Hassan:2013pca},
\be
\label{BiMbach} B_{\mu\nu} +
\mathcal{O}\left(\tfrac{P^3}{\mu^6}\right) =0\,.  
\ee 
At the lowest order, this is the equation of motion for conformal
gravity, $B_{\mu\nu}=0$, which is invariant under Weyl scalings
$\gmn\rightarrow e^{\phi(x)}\gmn$, since $B_{\mu\nu} \rightarrow
e^{-\phi(x)}B_{\mu\nu}$. Hence, at the lowest order in a curvature
expansion, the equations of the bimetric model \eqref{Pmpar} share the
Weyl symmetry of conformal gravity. The corresponding transformation
of $\fmn$ can be obtained from \eqref{fsol}. The question is if this
is just an accidental symmetry of the bimetric equations at the
four-derivative level, or if the Weyl scaling could be corrected by
adding higher-derivative terms to maintain the symmetry at higher
orders in the curvature expansion, thereby indicating a gauge symmetry
of the full bimetric equations. In this paper we investigate this
question systematically and, in section \ref{sec:sym}, show that the
corrections to Weyl scaling can be computed at least up to
six-derivative terms, equivalent to adding up to 10-derivative
corrections to the Bach equation. Before that, in the next section, we 
consider the relation to partial masslessness and conformal gravity at
the quadratic level.   

\section{Linear analysis: PM theory, conformal gravity, and the
  bimetric model}\label{sec:rel}

The bimetric model specified by \eqref{Pmpar} is precisely the one
identified in \cite{Hassan:2012gz} based on an analysis of partially
massless (PM) gauge symmetry in the linearised theory. It propagates
six modes (instead of the generic seven) around Einstein
backgrounds. The question is if it can lead to a better understanding  
of partial masslessness at the nonlinear level.

Traditionally, PM symmetry is studied in the context of the linear
Fierz-Pauli (FP) theory. Attempts to find nonlinear generalisations
are also often influenced by the FP setup, for example, in looking for
a {\it fundamental} PM field, or in modelling the PM transformations
of the nonlinear fields after the linear theory. On the other hand, it
is also known that conformal gravity exhibits a spectrum similar to
the linear PM theory, except that now, the underlying gauge symmetry
is Weyl invariance instead of PM symmetry, and the theory has a spin-2
ghost. After a brief review of these issues, in this section we
consider the bimetric model at the quadratic level and show that it
provides a unified description of linear PM theory as well as
linearised conformal gravity. We argue that the bimetric setup
provides a more powerful framework for finding a nonlinear
generalisation of PM theory.

\subsection{Partial masslessness in the Fierz-Pauli framework}
\label{sec:PMFP}

The Fierz-Pauli equation for a massive spin-2 field
$\delta\mmn$ in a de Sitter metric $\bgmn$ is, 
\begin{align}\label{FP}
\mathcal{E}_{\mu\nu}^{\rho\sigma}\delta M_{\rho\sigma} 
-\Lambda (\delta \mmn-\tfrac{1}{2}\bgmn\delta M)
 +\tfrac{1}{2}m^2_{\mathrm{FP}} (\delta\mmn-\bgmn\delta M)=0\,,
\end{align}
where the linearised Einstein operator is given by,
\begin{align}\label{kinop} 
(\mathcal{E}\delta M)_{\mu\nu}
\equiv -\tfrac{1}{2}\big(
\delta^\rho_\mu\delta^\sigma_\nu\bar{\nabla}^2+\bar{g}^{\rho\sigma}
\bar{\nabla}_\mu\bar{\nabla}_\nu-\delta^\rho_\mu\bar{\nabla}^\sigma
\bar{\nabla}_\nu &-\delta^\rho_\nu\bar{\nabla}^\sigma\bar{\nabla}_\mu
\nn\\ 
&-\bar{g}_{\mu\nu}\bar{g}^{\rho\sigma}\bar{\nabla}^2+\bar{g}_{\mu\nu}
\bar{\nabla}^\rho\bar{\nabla}^\sigma\big)\delta M_{\rho\sigma}\,.
\end{align}
It is well known that when the Higuchi bound is satisfied
\cite{Higuchi:1986py}
\beqn\label{Higuchi}
m^2_\mathrm{FP}=\tfrac{2}{3}\Lambda,
\eeqn
the FP equation becomes invariant under gauge transformations 
\cite{Deser:2001pe},  
\beqn\label{PM}
\Delta(\delta\mmn)=(\nabla_\mu\nabla_\nu+\tfrac{\Lambda}{3}\bgmn)\,\xi(x)\,.
\eeqn
Consequently, the linear spin-2 field $\delta\mmn$ has four
propagating modes, instead of the usual five for a massive
field. This is related to the fact that the de Sitter group also has a 
four-component ``partially massless'' representation, besides the
usual five-component massive representation \cite{Deser:1983mm,
  Deser:2001pe, Deser:2001us}. To take gravity into account, we may
add a massless spin-2 field $\delta G_{\mu\nu}$ with two propagating
modes, to the linear PM theory.   

An interesting question, investigated by many authors
\cite{deRham:2012kf, Deser, Hassan:2012gz, Hassan:2012rq,
  Deser:2013uy, deRham:2013wv, Hassan:2013pca,
  Deser:2013gpa,Joung:2014aba, Garcia-Saenz:2014cwa,
  Alexandrov:2014oda}, is if the linear PM theory in de Sitter
background can be extended to arbitrary backgrounds and if it can be
generalised to a nonlinear theory in a background independent
way. However, some features of equation \eqref{FP} may be taken as 
indications that the FP setup is not an adequate starting point for
such generalisations: 
\begin{enumerate}
\item In the FP setup, there is no analogue of PM theory in flat
  spacetime, indicating a preferred choice of background. For
  $\Lambda=0$, \eqref{FP} and \eqref{Higuchi} describe a massless
  spin-2 field with two propagating modes. In this case, $\delta\mmn$
  cannot be a PM field since, unlike the de Sitter group, the Poincar\'{e}
  group has no four-component spin-2 representation.
\item In a theory with general covariance, $\delta\mmn$ cannot be a
  fluctuation of some {\it background independent} tensor field
  $\mmn$. This is because \eqref{FP} is not invariant under
  infinitesimal reparameterisations of $\delta\mmn$. Then, $\delta\mmn$
  must the fluctuation of some background dependent (and hence, not
  fundamental) tensor field $M_{\mu\nu}$ in such a way that, precisely
  in de Sitter spacetimes, it becomes reparameterisation
  invariant.\footnote{This leaves out the possibility that
    $\delta\mmn$ is a nonlinear field with some nonlinear completion
    of \eqref{FP}.} Below we show that these issues can be naturally
  addressed in a bimetric setup.
\end{enumerate}

\subsection{Linearised conformal gravity and PM in bimetric framework}
\label{sec:lindS}

Conformal gravity (CG) is defined by the action $\int \dd^4x\,
\sqrt{g}~W^2$, where $W_{\mu\nu\rho\sigma}$ is the Weyl curvature
tensor. The action is invariant under Weyl scalings of the metric
$g\rightarrow e^\phi g$. The equation of motion is the Bach equation
$B_{\mu\nu}=0$. Due to Weyl invariance, this equation propagates six
modes of which two are healthy and four are ghosts (or {\it vice
  versa}, depending on the overall sign of the action). As observed in
\cite{Maldacena:2011mk}, in a de Sitter background, the linear
spectrum is similar to the spectrum of the PM theory with an extra
massless spin-2 field, except that one of the fields is now a ghost,
and the PM symmetry is replaced by Weyl invariance. This may be taken
as a hint of a connection between CG and PM theories. An attempt to
identify a nonlinear PM field in CG was made in \cite{Deser} but no
such field was found. Note that unlike the PM theory in the FP
framework, conformal gravity admits flat space as a background around
which it propagates six modes; two massless spin-2 fields and one
massless vector \cite{Riegert:1984hf}.

Now, we consider the bimetric model \eqref{Pmpar} at the quadratic
level and show that it provides a unified description of both linear
PM theory and linearised conformal gravity.  The Einstein-de Sitter
solutions in this model are of the type $\bfmn=c^2\bgmn$ where the
equations leave the constant $c^2$ arbitrary. The arbitrariness
is unique to this model and is a consequence of PM symmetry (see
footnote \ref{c^2} below). In these backgrounds, the fluctuations 
$\delta\gmn=\gmn-\bgmn$ and $\delta\fmn=\fmn-\bfmn$ combine into
massless and massive spin-2 modes~\cite{Hassan:2012wr}, \be \delta
\ggmn=\delta \gmn +\alpha^2\delta \fmn\,,\qquad \delta\mmn=\delta
\fmn-c^2\delta\gmn\,.
\label{dGdM}
\ee
The cosmological constant and Fierz-Pauli mass in this model 
satisfy the Higuchi bound and are given by (in the background metric
$\bgmn$ and using the notation in \eqref{mualpha}),  
\beqn\label{ccmfp}
\Lambda_g=\tfrac{3}{2}m^2_\mathrm{FP}=
3\beta_2\,\mu^2   (\alpha^{-2}+c^2)\,.
\eeqn
The linearised bimetric action, diagonalised into the above mass
eigenstates is,   
\begin{align}\label{linbim}
S_\mathrm{lin}=&\tfrac{-m_g^{2}}{1+\alpha^2c^2} \int\dd^4 x \,\Big[
  \delta\ggmn \mathcal{E}^{\mu\nu\rho\sigma}
  \delta G_{\rho\sigma} -\tfrac{\Lambda_g}{2}(
  \delta G^{\mu\nu}\delta \ggmn-\tfrac{1}{2}\delta G^2 )\\ 
&+\alpha^2c^{-2}\Big\{\delta\mmn \mathcal{E}^{\mu\nu\rho\sigma}
\delta M_{\rho\sigma} -\tfrac{\Lambda_g}{2}
(\delta M^{\mu\nu}\delta \mmn-\tfrac{1}{2}\delta M^2 )
 +\tfrac{\Lambda_g}{6} (\delta M^{\mu\nu}
\delta \mmn-\delta M^2 ) \Big\} \Big]\,, \nn
\end{align}
where $\mathcal{E}$ is the linear Einstein operator defined in
\eqref{kinop}.
This action is invariant under linearised diffeomorphisms of
$\delta G_{\mu\nu}$ as well as the linear PM gauge transformation,
\be
\label{Mpm}
\Delta(\delta\mmn)=A (\bar\nabla_\mu\partial_\nu+\tfrac{\Lambda_g}{3}
\bgmn)\xi(x),
\ee
with gauge parameter $\xi(x)$ and for any $A$. In principle, this may 
also be accompanied by a (restricted) coordinate transformation of
$\delta G_{\mu\nu}$,   
\beqn\label{Gpm}
\Delta(\delta G_{\mu\nu})=B\,\bar\nabla_\mu\partial_\nu\xi(x)\,.
\eeqn
Using \eqref{dGdM} one can easily work out the corresponding
transformations for the original variables $\Delta(\delta\gmn)$ and
$\Delta(\delta\fmn)$.\footnote{\label{c^2}As shown in
  \cite{Hassan:2012gz}, for constant $\xi$, these transformations can
  be integrated to finite ones only if $c^2$ is undetermined. This, in
  turn, uniquely led to the parameters \eqref{Pmpar}. Thus, changes in
  $c^2$ are associated with constant PM ``gauge" transformations
  restricted to the proportional backgrounds. For more general
  cosmological backgrounds, the equations leave a time-dependent
  function undetermined \cite{Hassan:2012gz}.}  Either $A$ or $B$ can
be absorbed in a rescaling of $\xi$, as long as it remains
non-singular. Beyond that, the arbitrariness in $A$ and $B$ cannot be
fixed in the quadratic theory. In section \ref{reltoPM} we show one
can set $B=1$ and nonlinear considerations in bimetric theory 
determine,  
\be
A=\tfrac{1}{2\alpha^2}(1-\alpha^2c^2)\,.  
\ee 
Owing to the above gauge symmetries, $\delta\mmn$ is a partially
massless field and $\delta G_{\mu\nu}$ is a massless field, with four
and two propagating modes, respectively. For positive values of $c^2$,
non of the modes is a ghost and (\ref{linbim}) is a healthy action.

As pointed out in \cite{Maldacena:2011mk}, conformal gravity in a de Sitter
background has the same structure of modes as above, except that the
PM field is a ghost. The reason for this similarity can be easily
understood at the quadratic level. For negative values of $c^2$ in the
action \eqref{linbim} the field $\delta\mmn$ becomes a ghost because
the sign of its kinetic term changes. In this case the linearised
bimetric action can be related to the linearised conformal gravity
action. To see this, consider $0>c^2> -\alpha^2$, and rescale the
fluctuations by real constants (with $|c|=\sqrt{{-c^2}}$),  
\be\label{scalings} 
\delta\mmn\longrightarrow\sqrt{\tfrac{6}{\Lambda_g}}\,
\frac{|c|}{\alpha m_g} \delta\mmn\,,\qquad
\delta\ggmn\longrightarrow\sqrt{\tfrac{\Lambda_g}{6}}\,
\frac{1}{m_g}\delta\ggmn\,,
\ee 
followed by an additional field redefinition,
\beqn\label{redefg}
\delta\ggmn'=\delta\ggmn-\tfrac{6}{\Lambda_g}\delta\mmn\,.
\eeqn
Only for $c^2<0$, this replaces the kinetic term of $\delta\mmn$ by a
kinetic mixing term,
\begin{align}
\label{linBMaux}
S'_\mathrm{lin}=&\tfrac{1}{1+\alpha^2c^2}\int\dd^4 x \,\Big[ 
\tfrac{\Lambda_g}{6}\Big(-\delta\ggmn'
\mathcal{E}^{\mu\nu\rho\sigma}\delta G'_{\rho\sigma}
+\tfrac{\Lambda_g}{2}\left(\delta G'^{\mu\nu}\delta \ggmn'-
\tfrac{1}{2}\delta G'^2\right)\Big)\nn\\
&-2\delta\mmn \mathcal{E}^{\mu\nu\rho\sigma}\delta G'_{\rho\sigma}
+\Lambda_g\left(\delta M^{\mu\nu}\delta \ggmn'-\tfrac{1}{2}
\delta M\delta G'\right)+\delta M^{\mu\nu}\delta \mmn-\delta M^2
\Big]\,,
\end{align}
This action is the linearised form of the auxiliary-field formulation
of conformal gravity \cite{Kaku:1977pa},  
\begin{align}\label{linca}
S_\mathrm{aux}=\tfrac{1}{1+\alpha^2c^2}\int\dd^4x\sqrt{G'}\,\Big[
\tfrac{\Lambda_g}{6}\Big(R-2\Lambda_g\Big)-M^{\mu\nu}\Big(
\mathcal{G}_{\mu\nu}+\Lambda_g\ggmn'\Big)+M^{\mu\nu}M_{\mu\nu}-M^2\Big],
\end{align}
where $R$ and $\mathcal{G}_{\mu\nu}$ are the scalar curvature and 
Einstein tensor of $\ggmn'$. The solution to the $\mmn$ equation of
motion,  
\be
\mmn=-\tfrac{\Lambda_g}{6}\ggmn'+\tfrac{1}{2}\left(R_{\mu\nu}-\tfrac{1}{6}
\ggmn' R \right)\,,
\ee
when plugged back into (\ref{linca}), gives the conformal gravity
action for $\ggmn'$,  
\be
S_\mathrm{C}=-\tfrac{1}{4(1+\alpha^2c^2)}\int\dd^4 x \sqrt{G'}\,
\left(R^{\mu\nu}R_{\mu\nu}-\tfrac{1}{3}R^2\right)\,.
\ee
It is then obvious that integrating out $\delta M_{\mu\nu}$ in
\eqref{linBMaux} leads to the linearised form of this action. We
emphasise that the manipulations \eqref{scalings} and \eqref{redefg},
remove the $\delta\mmn$ kinetic term only if $c^2<0$. Thus,
considering the bimetric model in a regime with spin-2 ghosts is
crucial for making the connection to conformal gravity at the level of
the action.

We conclude that bimetric theory, around its proportional backgrounds,
gives a unified description of linearised conformal gravity and linear
PM theory. At the quadratic level, both these theories are different
phases of the same bimetric model. It should be emphasised that
although the undetermined modulus parameter $c^2$ is equivalent to a
gauge parameter, it is not possible to start with $c^2>0$ and reach
$c^2<0$ by continuous gauge transformations. This is because such
transformations would necessarily have to cross $c^2=0$, for which one
of the metrics becomes singular. Hence, the ghost-free linearised PM
theory does not lie on the same gauge orbit as the linearised
conformal gravity.

As a final remark, the quadratic bimetric and CG actions are 
equivalent only for $c^2<0$, while for $c^2>0$, the actions are 
not equivalent. However, for both $c^2>0$ and $c^2<0$, the linearised  
bimetric equations lead to the linearised Bach equation. For any
$c^2$, the linearised bimetric equations are, 
\begin{align}
\mathcal{E}_{\mu\nu}^{~~\rho\sigma}\delta g_{\rho\sigma}-\tfrac{\Lambda_g}{2}
\left(\delta \gmn-\tfrac{1}{2}\delta g\,\gmn\right)
-\tfrac{2\alpha^2\Lambda_g}{3(1+\alpha^2c^2)}\Big(\delta\fmn- 
c^2\delta\gmn-(\delta f-c^2\delta g)\gmn\Big)&=0\,,\\
\mathcal{E}_{\mu\nu}^{~~\rho\sigma}\delta f_{\rho\sigma}-
\tfrac{\Lambda_g}{2} \left(\delta \fmn-\tfrac{1}{2}\delta f\,
\gmn\right) +\tfrac{2\Lambda_g}{3(1+\alpha^2c^2)}\Big(\delta\fmn
-c^2\delta\gmn-(\delta f-c^2\delta g)\gmn\Big)&=0\,.
\end{align}
Solving the first of these equations for $\delta\fmn$ and plugging the
result into the second always gives the linearised Bach equation for
$\delta\gmn$,  
\beqn
\frac{\delta B_{\mu\nu}}{\delta g_{\alpha\beta}}\,\delta g_{\alpha\beta}=0\,.
\eeqn

\subsection{Linear theory in flat space}\label{sec:PMflat}

The discussion in the previous subsection is valid only for
$\Lambda_g\neq 0$. However, unlike the PM theory in the FP formulation
\eqref{FP}, the bimetric model also has a flat space solution with six 
propagating modes around it, without a conflict with the Poincar\'e
group representations. From the expression for $\Lambda_g$
\eqref{ccmfp} it is evident that the flat space solution corresponds
to a \textit{choice} of 
\be 
c^2=-\alpha^{-2}\,. 
\label{flat}
\ee 
As mentioned above, this value cannot be reached by continuous 
transformations from $c^2>0$.\footnote{Flat space solutions do
  not exist in the corresponding massive gravity model
  of~\cite{deRham:2013wv}. In this case, the cosmological constant of
  the proportional backgrounds is $\Lambda_g=3c^2 \mu^2\beta_2$, where
  again $c$ is undetermined. Implementing $\Lambda_g=0$ by setting
  $c^2=0$ leads to a singular metric (since $\fmn$ is
  fixed). More generally, in the massive gravity limit, the CG-related
  phase of the bimetric model as well as many other solutions
  disappear \cite{Hassan:2014vja}, although more pathological
  solutions for $c^2<0$ could survive.}  

With the choice \eqref{flat}, the expressions of the massless and
massive modes in (\ref{dGdM}) coincide or, equivalently, the
expressions for $\delta\gmn$ and $\delta \fmn$ in terms of the mass
eigenstates become singular. It is therefore not possible to
diagonalise the equations in terms of spin-2 mass eigenstates. These
are no longer the correct variables to work with, and the earlier
discussion of the relation to conformal gravity, formulated in terms
of $\delta G_{\mu\nu}$ and $\delta M_{\mu\nu}$, is not valid.  This
also resolves the conflict with the absence of spin-2 PM
representations of the Poincar\'{e} group. The equations are, however,
well-defined in terms of the original $\delta\gmn$ and $\delta\fmn$
and become most transparent when expressed in terms of $\delta
g_{\mu\nu}$ and $\delta M_{\mu\nu}=\delta G_{\mu\nu}=\delta
g_{\mu\nu}+\alpha^2\delta f_{\mu\nu}$,
\begin{subequations}
\label{pmflateq}
\begin{align}
{\mathcal{E}'}^{\rho\sigma}_{\mu\nu}\delta M_{\rho\sigma}&=0\,,
\label{pmflateq1} \\
{\mathcal{E}'}^{\rho\sigma}_{\mu\nu}\delta g_{\rho\sigma}+\alpha^{-2}
\mu^2\beta_2 \Big(\delta M_{\mu\nu}-\eta_{\mu\nu}\delta
M_\rho^\rho\Big) &=0\,,
\label{pmflateq2}
\end{align}
\end{subequations}
where ${\mathcal{E}'}^{\rho\sigma}_{\mu\nu}$ is the linearised
Einstein operator \eqref{kinop} in flat space.  Despite the form of
the first equation, $\delta M_{\mu\nu}$ is not a massless spin-2 field
because the entire system is not invariant under linearised coordinate
transformations of $\delta M_{\mu\nu}$. Coordinate transformations of
$\delta g_{\mu\nu}$,   
\beqn
\delta g_{\mu\nu}\longrightarrow\delta g_{\mu\nu}+\partial_\mu\xi_\nu+
\partial_\nu\xi_\mu\,,
\eeqn
are symmetries of~(\ref{pmflateq}) but $\delta g_{\mu\nu}$ does not
satisfy the equation of a massless spin-2 field. The system of
perturbation equations is also invariant under a scalar gauge symmetry
which transforms both~$\delta g_{\mu\nu}$ and~$\delta M_{\mu\nu}$, 
\beqn
\Delta\delta\gmn=\tfrac{\mu^2\beta_2}{\alpha^{2}}\xi\,\emn\,,\qquad
\Delta\delta M_{\mu\nu}=\partial_\mu\partial_\nu\xi\,.
\eeqn
Moreover, taking the divergence of (\ref{pmflateq2}) shows that
$\delta M_{\mu\nu}$ is transverse. Therefore, the number of
propagating degrees of freedom around flat background is  
\beqn
20~\text{components}~ &-& ~2\cdot4~\text{coordinate~transformations}
\nn\\
&-&~2\cdot1~\text{scalar~gauge~transformations}~-~4~
\text{constraints}~=~6\,.   
\eeqn
This is the same number of modes as found around a de Sitter
background, where the fluctuations are diagonalisable into a massless
and a partially massless field. Around flat space, however, no
interpretation in terms of mass eigenstates with spin-2 alone exist.  
In fact, the spectrum is the same as that of conformal gravity which
consists of two massless spin-2 and one massless spin-1 fields
\cite{Stelle:1976gc, Ferrara:1977mv, Fradkin:1981iu,
  Riegert:1984hf}. Depending on the overall sign of the action, four 
or two out of the six propagating modes are ghosts. 

\section{A systematic study of the extended Weyl symmetry}\label{sec:sym}

Motivated by the appearance of the Weyl invariant Bach equation at
lowest order in the derivative expansion of bimetric equations
(\ref{BiMbach}), we now investigate the possibility of extending this
invariance to higher-derivative terms in the expansion. 

\subsection{General form of symmetry transformations}
\label{criteria}
Let us consider the bimetric model \eqref{Pmpar} in more
detail. The action is given by \eqref{act} with the potential, 
\be
V=3\beta_2\left(\alpha^{-2}+\tfrac{1}{6}([S^2]-[S]^2)+\alpha^2\det{S}
\right), 
\label{Vpm}
\ee
where $[S]\equiv \Tr\,{S}$ and $S=\sqrt{g^{-1}f}$. The $g$ and $f$
equations of motion are,
\bsubeq
\label{pm-eom}
\begin{align}
&\tfrac{1}{\mu^2\beta_2}R^\mu_{~\nu}(g)-\tfrac{3}{\alpha^2}\delta^\mu_{~\nu}+ 
(S^2-[S]S)^\mu_{~\nu}=0\,, 
\label{pmg-eom}\\
&\tfrac{1}{\mu^2\beta_2}R^\mu_{~\nu}(f)-3\delta^\mu_{~\nu}+\tfrac{1}{\alpha^2} 
(S^{-2}-[S^{-1}]S^{-1})^\mu_{~\nu}=0\,,
\label{pmf-eom}
\end{align}
\esubeq
in which we have raised the first index with $g^{\mu\nu}$ and
$f^{\mu\nu}$, respectively, and used the notation in \eqref{mualpha}.  
Note that in terms of rescaled variables,   
\be
\gmn'=\alpha^{-1}\gmn\,,\qquad \fmn'=\alpha\fmn\,,
\label{scale}
\ee
the equations \eqref{pm-eom} take a very symmetric form, 
\bsubeq
\label{p-eom}
\begin{align}
&\hat\alpha R^\mu_\nu(g')-3\delta^\mu_\nu+(S'^2-[S']S')^\mu_\nu=0\,,  
\label{g'-eom}\\
&\hat\alpha R^\mu_\nu(f')-3\delta^\mu_\nu+(S'^{-2}-[S'^{-1}]
S'^{-1})^\mu_\nu =0\,,  
\label{f'-eom}
\end{align}
\esubeq
in which we have used the further notation (see \eqref{mualpha})
\be\label{defhata}
\hat\alpha \equiv\frac{\alpha}{\mu^2\beta_2}
\ee
These transform into each other under the interchange
$\gmn'\leftrightarrow \fmn'$, keeping all parameters fixed. This
property will play an important role in the considerations below.

Our aim is to check if we could find nontrivial, field-dependent
transformations $g'\rightarrow g'+\Delta g'(g', f')$ and
$f'\rightarrow f'+\Delta f'(g', f')$ (involving also a scalar gauge
parameter) that keep the above equations invariant. Since dealing with
the variation $\delta S$ of a square-root matrix is cumbersome, we do
not directly study the variations of \eqref{p-eom}, but rather attempt
to construct the symmetry perturbatively. First, we deduce some
general properties of such transformations. 

\paragraph{Interchange symmetry:} The $g'$-equation \eqref{g'-eom} can
be rendered invariant for any $\Delta g'(g',f')$ by choosing an
appropriate compensating transformation $\Delta f'(g',f')$. Such a
$\Delta f'$ can be computed for any $\Delta g'$ at least
perturbatively, for example, by using the perturbative solution for
$f'$ in terms of $g'$ and its curvatures in \eqref{fsol}. To find a
symmetry of both equations, one must be able to specify $\Delta g'$
such that the same variations also keep the $f'$-equation
\eqref{f'-eom} invariant. Let us assume that such transformations
exist.

On the other hand, since the $f'$-equation \eqref{f'-eom} is obtained 
from the $g'$-equation by simply interchanging $g'$ and $f'$, it must 
also inherit the symmetry of the $g'$-equation. In other words, the
$f'$-equation must also be invariant under,
\be
\Delta_\mathrm{new} g'= \Delta f'\vert_{f'\leftrightarrow g'}\,,\qquad  
\Delta_\mathrm{new} f'=\Delta g'\vert_{f'\leftrightarrow g'}\,.
\ee
A similar argument holds for a new symmetry of the $g'$-equation. But
if we know from various limits that the equations can at most admit
one symmetry, then the two sets of transformations must coincide, 
$\Delta g'\equiv\Delta_\mathrm{new} g'$, $\Delta f'\equiv
\Delta_\mathrm{new} f'$, that is,    
\be
\Delta g'=\Delta f'\vert_{f'\leftrightarrow g'}\,,\qquad  \Delta f'=
\Delta g'\vert_{f'\leftrightarrow g'}\,.
\label{condition}
\ee
This property will become very useful in constructing the symmetry
transformations. 

\paragraph{Field dependence convention:} The appearance of the
square-root matrix, $S=\sqrt{g^{-1}f}$, in the above equations makes
it difficult to investigate their symmetry directly by treating the
fields non-perturbatively (although still linearly in the symmetry
parameter). Hence here we will work with the derivative
expansion of the above equations by recasting the $g'$-and
$f'$-equations as the expansions \eqref{fsol} and \eqref{gsol} for
$f'$ and $g'$, respectively (see below). Then the gauge
transformations can also be expanded as,
\be
\Delta g' =\sum_{n=0} \Delta_{(2n)} g'\,,\qquad
\Delta f' =\sum_{n=0} \Delta_{(2n)} f'\,,
\ee
where the variation $\Delta_{(2n)}$ contains a total of $2n$
derivatives of the fields and the gauge parameter. In this approach,
the expression $\Delta g'(g',f')$ can be expanded in different ways
since a dependence on $f'$ can be re-expressed in terms of $g'$ using
\eqref{fsol}, and vice versa, thereby mixing different orders of
curvatures. This ambiguity can be avoided by following the convention
that $\Delta g'$ is expressed entirely in terms of $g'$ and its
curvatures, such that $\Delta g'=\Delta g'(g')$
  and, similarly, $\Delta f'=\Delta f'(f')$.

\paragraph{The bootstrap construction:} In the previous section we saw
that, on perturbatively eliminating $f'$ between equations
\eqref{p-eom}, one obtains the Bach equation for $g'$, plus
corrections involving higher powers of $R_{\mu\nu}(g')/\mu^2$. Thus,
at the lowest order, $\Delta_{(0)}\gmn'=\phi\gmn'$ is a symmetry. To
check if this symmetry extends to higher orders, a straightforward
approach would be compute the higher derivative corrections to the
Bach equation and see if corresponding higher order corrections
$\Delta_{(2n)}g'$ to the symmetry transformation exit.  This would
show the existence of an on-shell symmetry in this perturbative
framework. The variation $\Delta f'$ could then be computed by
plugging $\Delta g'$ in the expansion \eqref{fsol}  and expressing the
outcome entirely in terms of~$f'$ using \eqref{gsol}. The $\Delta g'$
and $\Delta f'$ obtained in this way will automatically satisfy
\eqref{condition}. In this approach, to compute, for example, the
4-derivative term in $\Delta g'$ one needs to know the 8-derivative
correction to the Bach equation, which is rather lengthy to compute
and manipulate.

Below we follow an alternative bootstrap approach where, starting with
$\Delta_{(0)}g'$, the transformation is systematically constructed
order by order, by using the perturbative expansions for $f'$ and $g'$
and imposing the interchange symmetry \eqref{condition}. The
transformations are then guaranteed to leave \eqref{p-eom} invariant
order by order.

\subsection{Structure of the derivative expansion of the bimetric
  equations} 

As discussed in section~\ref{sec:PM}, the $\gmn$ equation
\eqref{pmg-eom} contains $\fmn$ algebraically through $S^\mu_{~\nu}$
and hence can be solved perturbatively to express $\fmn$ in terms of
$R_{\mu\nu}(g)/\mu^2\beta_2$. A similar statement holds for the $\fmn$
equation \eqref{pmf-eom}.\footnote{\label{FS}In other words, equation 
  \eqref{g'-eom} can be viewed as algebraically determining $S'$ in
  terms of the matrix $\ha R^\mu_{~\nu}(g')-3 \delta^\mu_{~\nu}$, or 
  $\ha P^\mu_{~\nu}(g')-\delta^\mu_{~\nu}$, after subtracting
  $\tfrac{\ha}{6}R$. So, in principle, \eqref{g'-eom} can be
  re-expressed as,  
  \be 
  S'= F\left(\ha P^\mu_{~\nu}(g')-\delta^\mu_{~\nu}\right)\,, 
  \ee 
  where $F$ denotes some matrix function. Since $P^\mu_{~\nu}$
  commutes with $\delta^\mu_{~\nu}$,  for small enough $\ha
  P^\mu_{~\nu}$, $F$ can be expanded in a power series in the usual
  way. Alternatively, this power series can be obtained directly by
  perturbatively solving \eqref{g'-eom} for $S'$, which is our
  approach here. Finally we compute \eqref{symsolg} from $f=gS^2$.}   
For the model \eqref{Vpm}, the expressions (\ref{fsol}) for $\fmn$ and 
(\ref{gsol}) for $\gmn$ obtained in this way become,
\begin{subequations}
\label{pertsolpm}
\begin{align} 
\fmn&=-\tfrac1{\alpha^2}\gmn+\tfrac1{\beta_2\mu^2}\pmn+4\sum_{n=2}^\infty  
\tfrac{\alpha^{2n-2}}{(4\beta_2\mu^2)^n}\,\gamma_{\mu\nu}^{(2n)}[g]\,,
\label{pertsolpmf}
\\ 
\gmn&=-\alpha^2\fmn+\tfrac{\alpha^2}{\beta_2\mu^2}\tpmn+4\sum_{n=2}^\infty
\,\tfrac{\alpha^{2}}{(4\beta_2\mu^2)^n}\,\gamma_{\mu\nu}^{(2n)}[f]\,.
\label{pertsolpmg}
\end{align}
\end{subequations}
In terms of the rescaled variables \eqref{scale} and \eqref{defhata},
these take an interchange symmetric form,   
\begin{subequations}
\label{symsol}
\begin{align}
\fmnp&=-\gmnp+\ha\pmn+4\sum_{n=2}^\infty\left(\tfrac{\ha}{4}\right)^{n} 
\, \gamma_{\mu\nu}^{(2n)}[g']\,,\label{symsolg}\\
\gmnp&=-\fmnp+\ha\tpmn+4\sum_{n=2}^\infty\left(\tfrac{\ha}{4}\right)^{n}
\,\gamma_{\mu\nu}^{(2n)}[f']\,.\label{symsolf}
\end{align}
\end{subequations}
Here $\gamma^{(2n)}_{\mu\nu}[g]$ denote terms with $2n$ derivatives,
involving $n$ powers of $\pmn$ and $(n-1)$ powers of the inverse
metric $g^{\mu\nu}$ to contract the indices. In appendix
\ref{curvapp}, we provide their explicit form for terms up to fifth
order in $\pmn$, i.e.~up to $n=5$. The same functions
$\gamma^{(2n)}_{\mu\nu}$ appear in both equations due to the
$g'\leftrightarrow f'$ interchange symmetry of the model. Starting
from \eqref{p-eom}, one can easily write down recurrence relations
that generate $\gamma^{(2n)}$ to any order. Note that such expansions
are not valid in the massive gravity limit ($\alpha\rightarrow\infty$)
of bimetric theory. 

Equations \eqref{symsol} are perturbatively equivalent to the
equations of motion \eqref{p-eom} in the algebraic sense (for the
purposes of symmetry arguments that involve algebraic manipulations
alone, it is not necessary that they are also equivalent as
differential equations). Also, at the lowest orders, the two
expansions are compatible.\footnote{The purpose of these expansions
  here is not to accurately compute perturbative corrections, in which
  case one has to ensure that the two curvature expansions are
  mutually compatible. The idea rather is that if equations
  \eqref{p-eom} admit a gauge symmetry, then the expanded equations
  \eqref{symsol} will exhibit a symmetry of the form described here,
  irrespective of the details of the compatibility of the two
  curvature expansions.}
Finally, we note that all $\gamma^{(2n)}_{\mu\nu}$ possess the following
important properties: 
\begin{itemize}
\item $\gamma^{(2n)}_{\mu\nu}$ vanishes identically on Einstein
  backgrounds for which $P_{\mu\nu}\propto\gmn$. 
\item Beyond that, $\gamma^{(2n)}_{\mu\nu}$ vanishes identically on the
  homogeneous and isotropic solutions of~\cite{vonStrauss:2011mq} for
  which, in the above model, $P_{\mu\nu}\propto(\gmn+\alpha^2\fmn)$. 
\item Linear perturbations of $\gamma^{(2n)}_{\mu\nu}$ around Einstein
  backgrounds also vanish identically, 
\be
\left.\frac{\delta\gamma^{(2n)}_{\mu\nu}}{\delta g_{\rho\sigma}}
\right|_{\mathrm{Einstein}} \delta g_{\rho\sigma} =0\,.
\label{dgammadg}
\ee
\end{itemize}
These statements can straightforwardly be proven by noting that for
the above type of backgrounds, the complete equations of motion imply
$P_{\mu\nu}=\beta_2\mu^2(\gmn+\alpha^2\fmn)$. But this is saturated by
the first two terms in the above expansions, hence the contributions
from higher terms must vanish.\footnote{Up to $n=5$, this can be
  verified directly from the explicit expressions provided in Appendix
  \ref{curvapp}.} This implies that there is no contribution from any
of the~$\gamma^{(2n)}_{\mu\nu}$ on these backgrounds. In the same way
one verifies the statement about the linear variations.

The above properties imply that for backgrounds of the type $\bar
f=c^2\bar g$, all $\gamma^{(2n)}$ vanish and the expansions in
\eqref{pertsolpm} or \eqref{symsol} become exact. At the lowest order,
$\bar P_{\mu\nu}=0$ and one gets $\bar f=-\bar g$. At the quadratic
order, $\bar P_{\mu\nu}=(\Lambda_g/3)\bgmn$ leads to $\bar f=c^2\bar
g$, with no higher derivative contributions (with $c^2$ and
$\Lambda_g$ related by \eqref{ccmfp}). At the end of this section we
will comment more on the validity regime of these expansions.

\subsection{Perturbative bootstrap construction of gauge symmetry}
\label{symconstr}  
   
Now, employing the criteria collected in section \ref{criteria}, we
find the condition under which a gauge symmetry could exist and
describe a procedure for constructing the transformations to any
order. We explicitly show that a gauge symmetry exists to, at least,
six orders in derivatives.

\paragraph{Zero-derivative terms:} 
Guided by the Weyl invariance of the Bach equation,
which appeared in the analysis of section~\ref{sec:PM}, we take the
lowest-order gauge transformation of \eqref{p-eom} to involve an
infinitesimal Weyl scaling of $g'$. Then, the $g'$ equation
\eqref{g'-eom} can always be rendered invariant by a compensating
transformation of $f'$ which can be easily computed perturbatively
using the expression \eqref{symsolg} for $f'$ in terms of
$g'$. To lowest order in derivatives, this gives,
\beqn
\Delta_{(0)}\gmnp=\phi\gmnp \,,\qquad
\Delta_{(0)}\fmnp=-\phi\gmnp \,.
\eeqn
Using (\ref{symsolf}), $\Delta_{(0)}\fmnp$ can be expressed in terms
of $\fmnp$ alone. Thus to zeroth order in derivatives, one gets,
\beqn
\Delta_{(0)}\gmnp=\phi\,\gmnp \,,\qquad
\Delta_{(0)}\fmnp=\phi\,\fmnp \,.
\label{delta0}
\eeqn
Obviously, the lowest-order terms in the transformations satisfy
the criterion of interchange symmetry
\eqref{condition} and hence must keep both equations \eqref{p-eom}
invariant to this order. It is trivial to see that on ignoring
derivative terms the interaction contributions in \eqref{p-eom} are
invariant under such transformations.\footnote{Of course, the
  interaction contributions in \eqref{p-eom} are invariant under
  $g'\rightarrow Ag'$, $f'\rightarrow Af'$, for any invertible matrix
  $A$, but only the Weyl part of this admits an extension to include
  derivative terms.}

\paragraph{Two-derivative terms:}
Now we describe in detail the construction of the two-derivative terms
in the transformation. The procedure generalises to higher orders
straightforwardly.
\begin{itemize}
\item {\it Step\,1:} The first step is to determine the structure of
  all two-derivative terms that are generated by the zero-derivative
  part of the transformations. Using $\Delta_{(0)}g'$ from
  \eqref{delta0} in (\ref{symsolg}) and retaining terms with up to two
  derivatives gives the following compensating transformation,
\beqn
\Delta_{(0)}\fmnp+\Delta_{(2)}\fmnp= -\phi \gmnp-
\ha\nabla_\mu\partial_\nu\phi+\hdots\,. 
\eeqn
The dots denote still missing two-derivative terms. The field dependence
ambiguity discussed in section \ref{criteria} is fixed by re-expressing
$\Delta f'$ entirely in terms of $f'$. Using the expression
\eqref{symsolf} for $g'$ in terms of $f'$, the above
transformation at the two-derivative level then becomes
$\phi\fmnp-\ha\phi \tilde{P}_{\mu\nu}-\ha\tilde
\nabla_\mu\partial_\nu\phi$.  Since the equations treat $g'$ and $f'$
on equal  footing, it follows that similar two-derivative terms  must
also appear in $\Delta_{(2)}g'$. Hence we take, 
\begin{align}
\Delta_{(0)}\gmnp+\Delta_{(2)}\gmnp=\phi\gmnp+a_1\phi P_{\mu\nu}+
a_2\nabla_\mu\partial_\nu \phi\,, \label{dgtwo}
\end{align}
with constants $a_1$, $a_2$ to be determined. Now, using this in
(\ref{symsolg}) to recompute $\Delta\fmnp$, and once again expressing
its $g'$-dependence in terms of $f'$ gives,
\begin{align}
\Delta_{(0)}\fmnp+\Delta_{(2)}\fmnp=\phi\fmnp-\left(a_1+\ha\right)
\phi \tilde{P}_{\mu\nu}-\left(a_2+\ha\right)\tilde\nabla_\mu\partial_\nu\phi.   
\label{dftwo}   
\end{align}
By construction, these leave \eqref{symsolg} invariant on-shell up to
second order in derivatives. No new two-derivative terms are generated.

\item {\it Step\,2:} To ensure that the same transformations also
  keep \eqref{symsolf} invariant, it is sufficient to impose the
  $g'\leftrightarrow f'$ interchange symmetry
  \eqref{condition}, as argued in section \ref{criteria}. Comparing
  \eqref{dgtwo} and \eqref{dftwo}, one sees that this requires
  $a_1=a_2=-\ha/2$. Hence the transformations that leave both equations in
  \eqref{symsol} or \eqref{p-eom} invariant at the two-derivative  
  level are,   
\begin{subequations}
\label{PMtd}
\begin{align}
\Delta_{(0)}\gmnp+\Delta_{(2)}\gmnp&=\phi\gmnp-\tfrac{\ha}{2}
\Big(\phi P_{\mu\nu}+\nabla_\mu\partial_\nu\phi\Big) \,,\\
\Delta_{(0)}\fmnp+\Delta_{(2)}\fmnp&=\phi\fmnp-\tfrac{\ha}{2}
\Big(\phi\tilde P_{\mu\nu}+\tilde\nabla_\mu\partial_\nu\phi\Big)\,.
\end{align} 
\end{subequations}
\end{itemize}
Let us briefly comment on some ambiguities at the two-derivative level
that are avoided by the above construction. One can check explicitly
that even for arbitrary $a_1$ and $a_2$, the transformations are a
symmetry of \eqref{symsol} at the two-derivative level (although this 
is fixed by the higher-derivative terms). This is due to the fact that
at the two-derivative level, a coordinate transformation $\delta
x^\mu=-a_2g'^{\mu\nu}\partial_\nu\phi$ induces variations $\delta\gmnp
= 2a_2\nabla_\mu\partial_\nu\phi$ and $\delta\fmnp=-2a_2\tilde\nabla_\mu
\partial_\nu\phi$. Also, to this order, the equations \eqref{symsol}
are unchanged under field redefinitions $g'\rightarrow g'+a_1 P$ and 
$f'\rightarrow f'-a_1\tilde P$. Hence, $a_1$ and $a_2$ are not genuine
parameters of the new symmetry. In this case, imposing the 
exchange symmetry \eqref{condition} not only guarantees the invariance
of \eqref{symsolf}, given that \eqref{symsolg} is invariant by
construction, but it also fixes the coordinate and field redefinition
ambiguities. 

\paragraph{Extension to higher orders:} The construction of the
two-derivative terms straightforwardly generalises to higher orders.
Two situations arise depending on the number of derivatives $(2n+2)$
for even or odd $n$. Given transformations $\Delta g'$ and $\Delta f'$
to $4m$-derivative order (odd $n$), they can always be extended to
$(4m+2)$-derivative order. But starting with $\Delta g'$ and $\Delta
f'$ to $(4m-2)$-derivative order (even $n$), an extension to
$4m$-derivative order exists only under certain conditions. These need
to be satisfied in order for a complete perturbative expression of the
gauge symmetry to exist. Remarkably, the four-derivative terms turn
out to meet this condition and hence we are able to prove the presence
of a gauge symmetry up to six orders in derivatives.

In order to demonstrate the above, let us try to generalise the
two-derivative construction to terms with arbitrary number of
derivatives in the spirit of a proof by induction. For this we assume
that $\sum_{k=0}^{n}\Delta_{(2k)}g'$ and
$\sum_{k=0}^{n}\Delta_{(2k)}f'$ are gauge symmetries of \eqref{symsol}
at the $2n$-derivative level and satisfy the interchange symmetry
\eqref{condition}. This implies that if we use
$\sum_{k=0}^{n}\Delta_{(2k)} g'$ in \eqref{symsolg} and express the
resulting variation for $f'$ entirely in terms of $f'$ using
\eqref{symsolf}, then we get $\sum_{k=0}^{n}\Delta_{(2k)}
f'$. Moreover, the interchange symmetry insures the invariance of
\eqref{symsolf}.

 We are now looking for transformations $\sum_{k=0}^{n+1}\Delta_{(2k)}
 g'$ and $\sum_{k=0}^{n+1}\Delta_{(2k)} f'$ that keep the equations
 invariant to next order. The new purely $(2n+2)$-derivative terms,
 $\Delta_{(2n+2)} g'$ and $\Delta_{(2n+2)} f'$ can be constructed
 following the two steps outlined earlier.

\begin{itemize}
\item{\it Step\,1:} We start by finding all relevant
  $(2n+2)$-derivative terms, i.e., those that are generated from lower
  orders. Using $\sum_{k=0}^{n}\Delta_{(2k)} g'$ in \eqref{symsolg} to
  compute the variation for $f'$ produces directly a set of
  $(2n+2)$-derivative terms in terms of $g'$.  Furthermore, converting
  all $g'$ into $f'$ using \eqref{symsolf} produces additional terms
  that will appear in $\Delta_{(2n+2)} f'$.\footnote{Note that all
    terms with lower number of derivatives are already accounted for
    in $\sum_{k=0}^{n}\Delta_{(2k)}f'$.} Motivated by the interchange
  symmetry \eqref{condition}, we then introduce all the corresponding
  terms with arbitrary coefficients into $\Delta_{(2n+2)}g'$ and
  recompute the form of $\Delta_{(2n+2)} f'$ from \eqref{symsolg}. As
  a representative of the $(2n+2)$-derivative terms in the
  transformations, let us consider, for instance, a term of the type
  $A\phi \tilde P^{n+1}_{\mu\nu}$ (for a constant $A$) which is
  generated in $\Delta f'$ on using $\Delta g'$ at order $2n$ in
  \eqref{symsolg}, and expressing the result in terms of $g'$. This
  term contains $n$ powers of $f'^{\mu\nu}$ to contract the
  indices. The corresponding term must be included in
  $\Delta_{(2n+2)}  g'$ with an arbitrary coefficient $B$,
\begin{align}
\Delta_{(2n+2)}\gmnp=B\phi P^{n+1}_{\mu\nu} +\cdots\,.
\label{B}
\end{align} 
Now when $\Delta_{(2n+2)}f'$ is recomputed from (\ref{symsolg}) with
$\Delta_{(2n+2)} g'$ included, one gets,  
\begin{align}
\Delta_{(2n+2)}\fmnp=\left((-1)^{n+1}B+A\right)
\phi\tilde P^{n+1}_{\mu\nu}+\cdots.
\end{align} 
The $A$-term is generated from lower orders, while the $B$-term arises
directly from \eqref{B}. The factor of $(-1)^{n+1}$ arises from the
replacement $\gmnp=-\fmnp+\hdots$ in $\phi P^{n+1}_{\mu\nu}$ (we pick
up a minus sign for each of the $n$ factors of $g'^{\mu\nu}$), along
with the overall minus sign for computing the transformation of $f'$
from the first term of \eqref{symsolg}.  By construction, these
transformations leave \eqref{symsolg} invariant.

\item{\it Step\,2:} The second step is to impose the interchange
  symmetry \eqref{condition} in order to insure the invariance of
\eqref{symsolf}. This yields the condition, for any single term,
\beqn\label{condcoeff}
B=(-1)^{n+1}B-A\,.
\eeqn
Clearly, this can be solved for $B$ only for even $n=2m$. This
means that we can impose the interchange symmetry guaranteeing the
invariance of both \eqref{symsolg} and \eqref{symsolf} only for terms
with $(4m+2)$ derivatives. For terms with odd $n=2m-1$, or $4m$
derivatives, the construction fails unless $A=0$. This must hold for
all types of terms generated at this order. If this condition is met,
i.e.~if all contributions generated from lower orders  cancel each
other out, then the coefficients $B$ are left arbitrary by the
construction at this order in the transformation, though they may be
fixed by  higher orders.  
\end{itemize}
We conclude that an on-shell gauge symmetry can be constructed 
provided that, at each $4m$-derivative level, the contributions
generated from lower orders vanish on-shell. (Dis)proving the
existence of a nonlinear gauge symmetry in the equations
\eqref{symsol} or \eqref{p-eom} reduces to checking whether this
condition is violated or not. 

\paragraph{Four- and six-derivative terms:} The above analysis 
shows that, in this approach, an obstruction first arises at the
four-derivative level. For a gauge symmetry to exist, it is necessary
that the variation $\Delta_{(0)}g'+\Delta_{(2)}g'$ used in
\eqref{symsolg} does not generate any four-derivative terms in $\Delta
f'$. This is not obvious at first sight. However, the outcome of a
rather lengthy calculation outlined in appendix \ref{appfd} is that
such four-derivative terms are indeed not generated by terms with
fewer derivatives (this can be easily verified for a constant gauge
parameter $\phi$). Therefore, the four-derivative terms in the
transformations remain arbitrary at this level. They may need to be
fixed when considering the eighth-order contributions, which again
need to vanish on-shell due to \eqref{condcoeff}.

The above construction also shows that, once a symmetry at the
four-derivative level is established, invariance at the six-derivative
level is guaranteed. Hence we are able to  perturbatively construct
gauge transformations of \eqref{symsol} up to sixth order in
derivatives,   
\begin{subequations}
\label{ExtWeyl}
\begin{align}
\Delta\gmnp&=\Delta_{(0)}\gmnp+\Delta_{(2)}\gmnp+\Delta_{(4)}\gmnp
+\Delta_{(6)}\gmnp+\hdots\,,\\ 
\Delta\fmnp&=\Delta_{(0)}\fmnp+\Delta_{(2)}\fmnp+\Delta_{(4)}\fmnp
+\Delta_{(6)}\fmnp+\hdots\,.
\end{align} 
\end{subequations}
The result for the zero- and two-derivative terms is given in
\eqref{PMtd}. Terms with four derivatives in the transformations are
completely arbitrary at this stage; any choice will leave the
equations invariant up to sixth order in derivatives. The
six-derivative terms, $\Delta_{(6)}g'$ and $\Delta_{(6)}f'$, are again
determined by the construction described above (but are expected to
depend on the choice of the four-derivative terms). The explicit
expressions are tedious, but the derivation is
straightforward.\footnote{To find these transformations using
  corrections to the Bach equation, we would have had to compute up to
  $10$-derivative corrections to the Bach equation.}  We emphasise
that these transformations are covariant and background independent to
the extent that the perturbative expansion is valid.

\subsection{Relation to the partially massless gauge symmetry}
\label{reltoPM} 

In section \ref{sec:lindS} we described how the linear PM theory is
easily embedded in a covariant bimetric setup. In this section, we
have constructed transformations \eqref{ExtWeyl}, at least up to 6
orders in derivatives, that keep the bimetric equations invariant. In
the perturbative setup employed, the transformations are simply
generalisations of the Weyl scalings of the metrics augmented by
higher derivative corrections. This construction required no input
from the PM properties of the quadratic theory.

It is now easy to verify that the linear PM transformations
\eqref{Mpm}, \eqref{Gpm} are recovered from \eqref{ExtWeyl} at the
two-derivative level. This also fixes the arbitrariness of the
quadratic theory. In terms of the unscaled metrics $g$ and $f$,  
the two-derivative variations \eqref{PMtd} can be rewritten as (after    
adding and subtracting $\Lambda_g$ contributions using \eqref{ccmfp}),    
\begin{subequations}
\label{PMtd-cc}
\begin{align}
\Delta\gmn &=\tfrac{1}{2}(1-\alpha^2c^2)\phi\gmn-  
\tfrac{\alpha^2}{2\mu^2\beta_2}\Big(\phi\left[P_{\mu\nu}-
\tfrac{\Lambda_g}{3}\gmn\right]+\nabla_\mu\partial_\nu\phi\Big) \,,\\
\Delta\fmn &=\tfrac{1}{2}(1-\alpha^{-2}c^{-2})\phi\fmn
-\tfrac{1}{2\mu^2\beta_2} \Big(\phi\left[\tilde P_{\mu\nu}
-\tfrac{\Lambda_g}{3c^2}\fmn\right]
+\tilde\nabla_\mu\partial_\nu\phi\Big)\,,
\end{align} 
\end{subequations}
where, higher derivative terms have been suppressed. Let us restrict the
nonlinear fields $\gmn$ and $\fmn$ to $\gmn=\bgmn+\delta\gmn$ and
$\fmn=\bfmn+\delta\fmn$. The fixed background satisfies
$\bfmn=c^2\bgmn$, so the transformation only affects the fluctuations,
$\Delta\gmn=\Delta(\delta\gmn)$ and $\Delta\fmn=\Delta(\delta\fmn)$. 
For such a background, $\bar P_{\mu\nu}=\bar{\tilde P}_{\mu\nu}=
(\Lambda_g/3)\bar g_{\mu\nu}$. Since the transformations are already
linear in the small gauge parameter $\phi$, only the 
background pieces contribute to the right-hand-sides of
\eqref{PMtd-cc}. Then, at two-derivative level, the symmetry
transformations reduce to,
\begin{subequations}
\begin{align}
\Delta(\delta\gmn)&=
\tfrac{1}{2} \left(1-\alpha^2c^2\right)
\phi\bar{g}_{\mu\nu}-\tfrac{\alpha^2}{2\beta_2\mu^2}\bar\nabla_\mu
\partial_\nu\phi \,,\\
\Delta(\delta\fmn)&=
-\tfrac{1}{2\alpha^2}\left(1-\alpha^2c^2\right)
\phi\bar{g}_{\mu\nu}-\tfrac{1}{2\beta_2\mu^2}\bar\nabla_\mu
\partial_\nu\phi\,.
\end{align} 
\end{subequations}
For the mass eigenstates \eqref{dGdM}, this implies, 
\begin{subequations}
\label{PMtd2}
\begin{align}
\Delta(\delta M_{\mu\nu})&=\tfrac{(\alpha^2c^2-1)}{2\mu^2\beta_2}\left(
\bar\nabla_\mu\bar\nabla_\nu +\tfrac{\Lambda_g} {3}\bgmn \right)\phi
 \,,\\
\Delta(\delta G_{\mu\nu})&=-\tfrac{\alpha^2}{\mu^2\beta_2}
\bar\nabla_\mu\bar\nabla_\nu\phi
\end{align} 
\end{subequations}
This recovers the standard PM transformation of the quadratic theory
in Einstein-de Sitter backgrounds, \eqref{Mpm}, along with a
coordinate transformation of \eqref{Gpm} (for
$\alpha^2\phi=-\mu^2\beta_2\xi$). It also fixes the ambiguity
present in the quadratic theory. Note that $\Delta\mmn$ is invariant
under infinitesimal coordinate transformations of $g$ and $f$. 

In this approach, it is obvious that the standard form of the PM
transformation in terms of $\delta\mmn$ is very specific to
Einstein-de Sitter backgrounds. Away from such backgrounds, the
fundamental fields are $\gmn$ and $\fmn$, and a fundamental PM field
does not exist. Furthermore, the nonlinear transformations induce no
linear variation of $\delta\mmn$ when $\alpha^2 c^2=1$. We will
comment on an implication of this in the next section.

\subsection{Limitations of the perturbative expansion} 

The perturbative expansions we have used are valid for small
$R_{\mu\nu}(g)/(\beta_2\mu^2)$, or small $\ha P_{\mu\nu}(g')$, where
$1/\beta_2\mu^2$ or $\ha$ are the small expansion parameters. To
connect to the linear PM theory, we considered the expansions around 
Einstein-de Sitter (EdS) solutions for which $\bar f'=\alpha^2c^2 \bar
g'$. But these backgrounds have a curvature scale $\Lambda_g'\sim
1/\ha$ (as implied by \eqref{ccmfp} in terms of rescaled variables).
Thus, in these cases, 
\be 
\ha P_{\mu\nu}(\bar g')=(1+\alpha^2 c^2)\bar g_{\mu\nu}' 
\ee 
is not suppressed by $\ha$ and is small only when $c^2$ is close to
$-\alpha^{-2}$. In spite of this, the standard PM results were
recovered for any $c^2$ from the first two terms in the expansions.
The reason is that although now the 4- and higher-derivative terms
$\gamma^{(2n)}_{\mu\nu}(\bar g')$ are not suppressed by $\ha$, they
nonetheless vanish on EdS backgrounds and, hence, are small in the
vicinity of it, for reasons having to do with the structure of
the equations. The linear variations of $\gamma^{(2n)}_{\mu\nu}$
around EdS solutions also vanish. But, now the expansion in 
$\gamma^{(2n)}_{\mu\nu}$ is not perturbative since terms with
different $n$ may make small, but comparable contributions. Hence the
expansions we have used are definitely 
reliable as perturbative expansions when $|1+\alpha^2c^2|<1$, in which
case the spectrum of perturbations is not unitary. However, if the
presence of a symmetry could be established to all orders in this
regime, this would also imply the invariance of the full equations of
motion.  For $c^2>0$, the status of the expansion is less 
obvious although the linear results around EdS solutions are easily
reproduced.  In any case, the possibility of extending the symmetry
beyond the four-derivative Bach equation clearly points to additional
structure within the nonlinear equations.

It is easy to rewrite equations \eqref{p-eom} such that they are valid
perturbative expansions around general proportional backgrounds, but
this is not useful for finding the transformations. In \eqref{p-eom}
we can simply rewrite $R_{\mu\nu}(g')$ as $(R_{\mu\nu}(g')-\Lambda'_g
\gmn')+\Lambda'_g\gmn'$ and expand $S'$ in powers of $\ha(R_{\mu\nu}(g')
-\Lambda'_g\gmn')$, or equivalently, in powers of $\ha{\cal
  P}_{\mu\nu}(g')=\ha(P_{\mu\nu}(g')-(\Lambda'_g/3)\gmn')$. In the
notation of footnote \ref{FS}, the equations imply that $S'$ must have 
the form, 
\be 
S'=F\left(\ha{\cal P}^\mu_{~\nu}+(\ha\tfrac{\Lambda'_g}{3}
-1)\delta^\mu_{~\nu}\right)= F\left(\ha{\cal P}^\mu_{~\nu}+
\alpha^2c^2\delta^\mu_{~\nu} \right)\,,
\ee 
where $F(X)$ is a matrix function of the matrix $X$ and traces of
powers of $X$.  Since ${\cal P}^\mu_{~\nu}$ commutes with
$\delta^\mu_{~\nu}$, for small enough $\ha$, this can be expanded in a
power series in the usual way. In practice, this is again achieved by
solving the equation perturbatively for $S'$. The two lowest-order
terms are just rewritings of the corresponding terms in
\eqref{symsol}, $\fmnp=\alpha^2c^2\gmnp+\ha{\cal
  P}_{\mu\nu}(g')+\cdots$. At higher orders, the $\gamma^{(2n)}$ are
now functions of $\cal P$, rather than $P$, and the coefficients are
$c$-dependent. Close to EdS backgrounds, $\ha{\cal P}_{\mu\nu}$ are
small and the expansion is perturbative. However, under a variation,
$\ha\Delta{\cal P}_{\mu\nu}=\ha\Delta
P_{\mu\nu}-(1+\alpha^2c^2)\Delta\gmn'$, note that $\Delta g'$ is not
suppressed by a small $\ha$. Hence higher terms contribute to the
variation of lower terms and the expansion is not useful for finding
the transformations order by order.

Alternatively, in \eqref{p-eom} we can rewrite $P_{\mu\nu}(g')$
as $(P_{\mu\nu}-\tfrac{1}{3}\Lambda'_g\bgmn')+ \tfrac{1}{3}
\Lambda'_g\bgmn'$. Then, 
\be 
S'=F\left(\ha\hat{\cal P}^\mu_{~\nu}+(1+\alpha^2 c^2)\,
g^{'\mu\lambda}\bar g_{\lambda\nu}'  -\delta^\mu_{~\nu}\right)\,.
\ee 
Now, taking $\Delta\bar g=0$, the variation of $\hat{\cal
  P}_{\mu\nu}=P_{\mu\nu}-\tfrac{1}{3}\Lambda'_g\bgmn'$ does not mix
terms of different order. But $\hat{\cal P}_{\mu\nu}$ does not commute
with $(g'^{-1}\bar g')^\mu_{~\nu}$ and the expansion is
involved. Hence there is no easy way of improving over the expansions
considered here.  

\section{Discussion}\label{sec:sum}
In this section we first discuss our results and then argue that the
bimetric model considered here avoids the counter arguments presented
so far for the absence of a gauge symmetry in the bimetric setup,
at least in their present form.  
\subsection{Summary and discussion of results}
In this paper we investigated a possible gauge symmetry of the
equations of motion of a particular bimetric model specified by
\eqref{Pmpar}. The equations contain the square-root matrix
$S=\sqrt{g^{-1}f}$ and the expression for the variation of $S$ in
terms of the variations $\Delta g$ and $\Delta f$ is complicated. We
avoid this complication by working with appropriate perturbative 
expansions of the equation.  

In a small curvature regime, we perturbatively eliminate one of the
metrics, say, $\fmn$ between the two equations to get a higher
derivative equation for $\gmn$. At the lowest order, instead of the
Einstein equation for $\gmn$, for this particular model one obtains
the 4-derivative Bach equation of conformal gravity which is invariant
under Weyl scalings of $\gmn$.  In a systematic treatment, we show
that the Weyl scalings of $\gmn$ and $\fmn$ can be corrected at least
up to 6-derivative terms to maintain the gauge symmetry at higher
orders. A brute force computation of the same transformations, by
demanding the invariance of the higher-derivative extension of the
Bach equation, would have required calculating up to 10-derivative
corrections to the Bach equation. Furthermore, we find a sufficiency
condition for extending the symmetry to higher orders: In our
approach, at orders $4m$ there are obstructions to the symmetry which 
must vanish. Then a symmetry at levels $4m+2$ is insured to exist.
Once an obstruction vanishes at a given order $4m$, the
transformations at that level remain completely arbitrary and will be
fixed at order $4(m+1)$.

The above results are completely independent of de Sitter backgrounds
and linear partial masslessness. Nevertheless, it turns out that in
Einstein-de Sitter spacetimes, the lowest-order terms in the
transformations do reduce to the well-known PM gauge transformations
for a composite PM field. We also considered the bimetric model at the
quadratic level and showed that it provides a unified description of
both linear PM theory as well as linearised conformal gravity,
including in flat spacetime. We argued that the bimetric setup
provides a natural and more promising framework for finding a
nonlinear PM theory, without a fundamental PM field, as compared to
the usual Fierz-Pauli framework.   

Our analysis of the transformations as extensions of Weyl scaling is
intimately connected to the $g'\leftrightarrow f'$ interchange
symmetry of the equations and breaks down in the massive gravity limit
of the theory. Also the equations breakdown for the relevant bimetric
theory in 3 dimensions. Such two-derivative theories do not exist in
other dimensions. Thus the analysis indicates the absence of such a
symmetry in $d=3$, as well as in massive gravity. 

If the 6-derivative transformation of the equations of motion found
here can be extended to all orders, this would imply a local gauge
invariance of the original equations. In the present approach going
beyond the 6-derivative order is cumbersome, so the analysis is not
conclusive. However, it is evident from our findings that the
equations in this bimetric model possess a lot of hidden structure.  A
possibility is that a gauge symmetry exists and a better understanding
of the structure of equations may enable one argue for its presence
and even find its closed form expression. The model would then
propagate six instead of seven degrees of freedom around any
background and would become a ghost-free replacement for conformal
gravity. In this case, imbedding the model in a larger setup should
make the symmetry manifest.

It is also possible that the equations do not have a gauge symmetry
and our construction cannot be extended to all orders. Nonetheless, the
results show that the bimetric framework is more powerful than the
Fierz-Pauli framework in searching for a nonlinear PM theory. It is
also possible that a gauge symmetry exists in expected completions of 
the model. It is obvious that the bimetric model is the spin-2
analogue of the Proca theory in curved background. Hence, just like
Proca theory, one may need additional degrees of freedom to obtain
the model through a Higgs-like mechanism. This or other extensions may
be better candidates for finding a new gauge symmetry. If such a nonlinear PM theory 
exists it would be very interesting to consider it as an alternative for conformal gravity 
in the proposal of~\cite{tHooft:2011aa, Hooft:2014daa}, which attempts to embed the Standard Model in a renormalisable 
theory of gravity.

Finally, the symmetry we have found (at 6-derivative level) is an
on-shell symmetry of the equations of motion, except for the quadratic
theory, where it is also an off-shell symmetry of the action. In terms
of the nonlinear fields, even the constant scalings of the metrics at
the zero derivative level are not symmetries of the action, whereas
they trivially keep the equations of motion invariant. Of course,
off-shell symmetries of the action always imply invariance of the
equations of motion, but the converse is not always true.

\subsection{Discussion of the counter-arguments}

Recently, some arguments have been made against the existence of a
scalar gauge symmetry in the bimetric model with parameters
(\ref{Pmpar}). We have shown that this model goes beyond other
similar constructions, in particular in displaying PM and conformal
gravity behaviours in different phases, but we do not make any claims
beyond what we have explicitly computed. It is also interesting to see  
that the model can evade the counter arguments presented so far and
could suggest ways of improving them. First of all, since the
transformations we construct are on-shell invariances, any off-shell
analysis at the level of the action must also remain valid on-shell to
be relevant. 

\paragraph{Massive gravity based arguments:} The main argument for the
absence of a gauge symmetry in the bimetric model is based on the
corresponding massive gravity limit. The massive gravity model,
obtained in the $\alpha\rightarrow\infty$ limit \cite{Baccetti:2012bk}
of the bimetric model \eqref{Pmpar} (for the unscaled variables
$g,f$), was studied in \cite{Deser:2013uy, Deser:2013gpa,
  deRham:2013wv}. It also admits proportional backgrounds
$g=c^{-2}\bar f$ with arbitrary $c^2$, where $\bar f$ is now a {\it
  fixed} Einstein-de Sitter metric with
$\Lambda_g=3\beta_2\mu^2c^2$. The linear theory around this background
is precisely the linear PM theory reviewed in section \ref{sec:PMFP}.
Arguments were presented in \cite{Deser:2013uy, deRham:2013wv,
  Deser:2013gpa} that, in this case, the linear PM symmetry does not
extend to the nonlinear theory.  It was also argued that since massive
gravity can be obtained as a limit of bimetric theory, the absence of
a PM symmetry in the massive gravity model implies the same for the
bimetric model \cite{Deser:2013uy, deRham:2013wv, Deser:2013gpa}.

This argument assumes that the bimetric gauge transformations have an
$\alpha\rightarrow\infty$ limit. The results in \cite{Hassan:2013pca}
and in the present paper indicate that this is not the case, as was
explained in \cite{Hassan:2014vja} in detail.  We summarise the key
points here. 

(i) Our derivation of the symmetry to sixth order heavily depended on
the $g'\leftrightarrow f'$ interchange symmetry of the equations,
which is destroyed in the massive gravity limit.

(ii) Our starting point was that the $g$-equation of motion can always
be rendered invariant by an appropriate transformation of $f$. In
massive gravity, $f$ is a fixed metric that cannot be varied (not even
within a class of solutions of the Einstein equation for $f$).

(iii) The expansion \eqref{pertsolpmf} does not exist in massive
gravity, since a term with $2n$ derivatives diverges as
$\alpha^{2n-2}$ in the limit and there are infinitely many terms with
increasing powers of $\alpha$. Hence the massive gravity equations are
not writable as the Bach equation plus corrections and the Weyl
invariance at the lowest-order in the derivative expansion is not
obtainable. Massive gravity therefore lacks the remarkable structure
that exists in the bimetric equations.

(iv) The gauge transformations $\Delta g$, $\Delta f$ obtained from
\eqref{pertsolpm} or \eqref{symsol} also involve an expansion powers
of $\alpha$ and hence diverge in the massive gravity limit. The
solutions that are obtained by applying these transformations also do
not exit in the massive gravity limit. This indicates that these gauge
transformations indeed disappear in the massive gravity limit. Of
course, these statements are based on a perturbative expansion which,
in principle, could sum up into a closed expression with a well
defined $\alpha\rightarrow\infty$ limit. But, as pointed out in
\cite{Hassan:2014vja}, solutions that are singular in the limit do
exist. The key point is that massive gravity is a limit of bimetric
theory around specific classes of solutions that need to be specified
before taking the limit (a massive gravity limit of the entire
bimetric theory does not exist). As discussed in
\cite{Hassan:2014vja}, for any solution with a well-defined massive
gravity limit, there exists another solution which becomes singular in
this limit.\footnote{An example is the flat bimetric solution in
  section \ref{sec:PMflat} that is singular in the massive gravity
  limit}

If now a gauge transformation in nonlinear bimetric theory connects
these two types of solutions, the perturbative expansion for the
transformations will contain an infinite number of increasing powers
of $\alpha$ (as is the case for the transformations constructed here),
then clearly it cannot survive the massive gravity limit and the
corresponding solutions are also absent after the limit has been
taken. The form of the transformations derived here suggests that,
generically, every gauge orbit contains one solution around which the
bimetric theory has a massive gravity limit and the rest of the gauge
orbit is invisible in the massive gravity limit.

(v) So far we have argued that if the bimetric model \eqref{Pmpar} has
a gauge symmetry of the type discussed here, the symmetry will not
survive in the massive gravity limit. But this argument must exclude
the PM symmetry of linearised massive gravity around EdS
backgrounds. Indeed, in our analysis, the terms that diverge as
$\alpha\rightarrow\infty$ are $\gamma^{(2n)}_{\mu\nu}$ for $n\geq
2$. As mentioned earlier, on EdS backgrounds, $P_{\mu\nu}(\bar
g)=(\Lambda_g/3)\bgmn$, both $\gamma^{(2n)}_{\mu\nu}(\bar g)$ as well
as their first variations vanish \eqref{dgammadg}. Hence in this case
the transformations do have a massive gravity limit, as should be the
case.

\paragraph{Symmetry algebra arguments:}

Some recent interesting works have studied the restrictions imposed on
nonlinear PM theories by the closure of the PM symmetry algebras
\cite{Joung:2014aba,Garcia-Saenz:2014cwa}. Reference
\cite{Joung:2014aba} considered a general action for a PM field 
$\phi_{\mu\nu}$ interacting with a metric $G_{\mu\nu}$ and analysed
the theory to cubic order in $\phi_{\mu\nu}$. At the lowest order,
$G_{\mu\nu}$ was assumed to be invariant and $\phi_{\mu\nu}$
transformed as a linear PM field, but now in a general background.  
$\phi_{\mu\nu}$-dependent corrections to the transformations were
determined from the invariance of the quadratic and cubic
$\phi_{\mu\nu}$ terms. On imposing the closure of the algebra it was
found that whenever the PM field had a positive kinetic term, the
cubic interactions were imaginary, while real cubic interactions were
associated with a ghost PM field (as in conformal gravity). This lead
to the conclusion that a unitary interacting theory with PM symmetry
does not exist. \cite{Joung:2014aba} also emphasised that the action
studied covered bimetric theory. If true, this would imply that the
symmetry, at the 6-derivative level, found here is an artefact of the
$c^2<0$ regime of bimetric theory and cannot be extended to higher
orders. However, besides the analysis being off-shell, there is a
potential caveat. To show that the PM interactions considered also 
covered bimetric theory, \cite{Joung:2014aba} used the relations,
\be
\label{Gphi}
G_{\mu\nu}=\gmn+ \fmn\,,\qquad \phi_{\mu\nu}=\gmn-\fmn\,.  \ee and
re-expressed the bimetric action (for $m_g=m_f$) in terms of
$G_{\mu\nu}$ and $\phi_{\mu\nu}$. Due to the symmetry $S(g,f)=S(f,g)$
there will be no terms linear in $\phi_{\mu\nu}$ (as assumed in
\cite{Joung:2014aba}), but also there will be no cubic $\phi_{\mu\nu}$
terms that were crucial in the analysis of
\cite{Joung:2014aba}. Moreover, the identification of a PM field in
bimetric theory is background dependent and the $\phi_{\mu\nu}$ in
\eqref{Gphi} can be identified with a PM field only around
proportional backgrounds with $c^2=1$. Then, since $\alpha=1$, from
the equations in section \ref{reltoPM} it is evident that around
proportional backgrounds, (corresponding to $\phi_{\mu\nu}=0$), the PM
field does not transform. Note that in this case, the quadratic theory
still has an accidental linear PM symmetry, but this does not follow
from the nonlinear PM symmetry. The vanishing of the transformation
for $c^2=1$ is completely consistent with the absence of the cubic
$\phi_{\mu\nu}$ interactions. Hence, at least using the
identifications \eqref{Gphi} advertised in \cite{Joung:2014aba}, the
analysis there is avoided by the bimetric model.

Another general study of PM algebra was carried out in
\cite{Garcia-Saenz:2014cwa}. The analysis involved only one dynamical
spin-2 field but, nevertheless, seems related to our work. The authors
considered general nonlinear extensions of the PM symmetry in dS and,
by demanding closure of the transformation algebra, found a unique
candidate transformation. The form of this symmetry, and the fact that
it had a non-trivial flat space limit, led the authors to the
conclusion that such an invariance could not be realised in any
two-derivative single-field Lagrangian. These results are fully
consistent with ours. In order to compare our setup to the
single-field analysis of \cite{Garcia-Saenz:2014cwa} we have to
consider the single-field higher derivative equation
\eqref{BiMbach}. We found a candidate symmetry of those equations up 
to six-derivative order (in the transformation), which furthermore
also had a flat space counterpart. It is obvious that the
single-field equation \eqref{BiMbach} does not derive from any
two-derivative single-field action.  In fact, as a two-derivative
theory they follow from the bimetric action containing both $\gmn$ and
$\fmn$ whereas, when viewed as a single-field theory, they are
higher-derivative equations which derive from a non-local action
obtained after properly integrating out $\fmn$.

Finally, we should mention the constraint analysis of
\cite{Alexandrov:2014oda} in the $d=3$ version of the present bimetric
model given in \cite{Hassan:2012rq}. The outcome, that the model in
3-dimensions does not posses PM symmetry is consistent with the fact
that the expansions in \eqref{symsol} do not exist in the $d=3$
theory. 

It is interesting to see if a more refined version of the above
analysis could make a more definitive statement about PM theory in the
bimetric framework.

\vspace{20pt}

\noindent

{\bf Acknowledgments:}
We thank Luis Apolo, Latham Boyle, Cedric Deffayet, Bo Sundborg, Andrew Waldron and
Nico Wintergerst for helpful discussions. The work of ASM is supported
by ERC grant no.~615203 under the FP7 and the Swiss National Science
Foundation through the NCCR SwissMAP.  The research of MvS has
received funding from the European Research Council under the European 
Community's Seventh Framework Programme (FP7/2007-2013 Grant Agreement
no. 307934).

\appendix

\section{Details of bimetric theory and its curvature
  expansion}\label{app:details} 

\subsection{Structure of action and equations of
  motion}\label{app:detbim} 
 
The elementary symmetric polynomials that enter the ghost-free
bimetric action can be defined through the following recursion
formula, 
\beqn
e_n(S)=\frac{1}{n}\sum_{k=0}^{n-1}(-1)^{k+n+1}
\mathrm{Tr}[S^{n-k}]e_k(S)\,,\qquad e_0(S)=1\,.
\eeqn
Note that $e_4(S)=\det(S)$ and $e_n(S)=0$ for $n>4$, if $S$ is a 
$(4\times4)$-matrix. Explicit expressions can easily be written down. 
The contributions from the interaction potential to the bimetric
equations of motions are given by \cite{Hassan:2011vm}, 
\beqn
V_{\mu\nu}&\equiv&-\frac{2}{\sqrt{g}}\frac{\partial (\sqrt{g}~V)}{\partial g^{\mu\nu}}=g_{\mu\rho}\sum_{n=0}^3 \beta_n[Y^{(n)}(S)]^\rho_{~\nu}\,,\nn\\
\tilde{V}_{\mu\nu}&\equiv&-\frac{2}{\sqrt{f}}\frac{\partial (\sqrt{g}~V)}{\partial f^{\mu\nu}}=f_{\mu\rho} \sum_{n=0}^3 \beta_{4-n}[Y^{(n)}(S^{-1})]^\rho_{~\nu}\,.
\eeqn
in which we have used the definitions,
\beqn
[Y^{(n)}(S)]^\rho_{~\nu}\equiv\sum_{k=0}^n(-1)^ke_k(S)\,[S^{n-k}]^\rho_{~\nu}\,.
\eeqn
Since the matrices $S^{-1}$ and $S$ with indices raised or lowered by
either of the two metrics are symmetric, the same holds for the
$[Y^{(n)}(S)]^\rho_{~\nu}$. Note also the following identity, 
\beqn\label{CH}
[Y^{(4)}(M)]^\rho_{~\nu}=0\,,
\eeqn
 which holds for any $4\times 4$ matrix $M$ and follows directly from
 the Cayley-Hamilton theorem.

\subsection{Obtaining the curvature expansions}\label{app:obtHC}

The explicit expression for th perturbative solution for $\fmn$
obtained from the $\gmn$ equation~(\ref{g-eom})
reads~\cite{Hassan:2013pca},      
\be
\label{fsolapp}
\fmn=a^2\gmn+\tfrac{2a^2}{s_1\mu^2}P_{\mu\nu}+\tfrac{a^2(s_1+2s_2)}
{s_1^3\mu^4}P^2_{\mu\nu}+\tfrac{2a^2s_2}{s_1^3\mu^4}\left[
\tfrac{1}{3}e_2(P)\gmn-PP_{\mu\nu}\right]+\mathcal{O}\left(
\tfrac{P^3}{\mu^6}\right)\,,
\ee 
where $\mu^2={m^4}/{m_g^2}$ and all indices on the right-hand side are
contracted with $\gmn$. The $s_n$ are particular combinations of
bimetric parameters given by,  
\be
s_n=\sum_{k=n}^3{3-n\choose k-n}a^k\beta_k\,.
\ee 
The value of the coefficient $a^2$ is obtained by solving the equation
$s_0=0$ for $a$ (where we only consider solutions that result in real
values for $a^2$). The solution for $\gmn$ obtained from the $\fmn$
equation~\eqref{f-eom} has the very similar form,     
\be
\label{gsolapp}
\gmn=\tilde{a}^2\fmn+\tfrac{2\tilde{a}^2}{\tilde{s}_1\tilde{\mu}^2}
\tilde{P}_{\mu\nu}+\tfrac{\tilde{a}^2(\tilde{s}_1+2\tilde{s}_2)}{\tilde{s}_1^3
\tilde{\mu}^4}\tilde{P}^2_{\mu\nu}+\tfrac{2\tilde{a}^2\tilde{s}_2}
{\tilde{s}_1^3\tilde{\mu}^4}\left[\tfrac{1}{3}e_2(\tilde{P})\fmn-
\tilde{P}\tilde{P}_{\mu\nu}\right]+\mathcal{O}\left(\tfrac{\tilde{P}^3}{\tilde{\mu}^6}
\right)\,,
\ee
where $\tilde{P}_{\mu\nu}\equiv\pmn(f)$ is the Schouten tensor for
$\fmn$. Indices are now contracted with $\fmn$ and, 
\beqn
\tilde{\mu}=\frac{m^4}{\alpha^2 m_g^2}\,,\qquad
\tilde{s}_n=\sum_{k=n}^3{3-n\choose k-n}\tilde{a}^k\beta_{4-k}\,,
\eeqn 
with $\tilde{a}$ being a solution to the polynomial equation
$\tilde{s}_0=0$.   

We can use the expression \eqref{fsolapp} for $\fmn$ to eliminate it
from its own equation. For instance, inserting (\ref{fsolapp}) into
the Einstein tensor for $\fmn$ results in,  
\be
\tilde{\mathcal{G}}_{\mu\nu}(f)={\mathcal{G}}_{\mu\nu}(g)-\tfrac{1}{s_1\mu^2}
\Big( \nabla^2P_{\mu\nu}&+&\nabla_\mu\nabla_\nu 
P-\nabla^\rho\nabla_\mu P_{\rho\nu}-\nabla^\rho\nabla_\nu P_{\rho\mu} \nn\\
&+&3PP_{\mu\nu}-\gmn\left[P^{\alpha\beta}P_{\alpha\beta}+\tfrac{1}{2}P^2
  \right] \Big)+\mathcal{O}\left(\tfrac{P^3}{\mu^4}\right)\,.
\ee
The contributions from the interaction potential read as, 
\begin{align}
\tfrac{\mu^2}{\alpha^2}\tilde{V}_{\mu\nu}&=\tfrac{\mu^2\Omega}{a^2\alpha^2} 
\gmn+\tfrac{1}{a^2\alpha^2}\mathcal{G}_{\mu\nu}+
\tfrac{2\Omega}{a^2\alpha^2s_1}P_{\mu\nu}\nn\\
&+\tfrac{1}{a^2\alpha^2s_1^3\mu^2}\left[x_1P^\rho_\mu P_{\rho\nu}+
x_2 PP_{\mu\nu}+\tfrac{1}{6}\gmn(x_3P^{\alpha\beta}P_{\alpha\beta}-x_2P^2)
\right]+\mathcal{O}\left(\tfrac{P^3}{\mu^4}\right)\,,
\end{align}
in which the coefficients $x_n$ are given by,
\begin{align}
	x_1=2s_1^2+\Omega(s_1+2s_2)\,,\qquad
	x_2=-3s_1^2-2s_2\Omega\,,\qquad
	x_3=3s_1^2-2s_2\Omega\,,
\end{align}
and we have defined,
\begin{align}
	\Omega=a\beta_1+3a^2\beta_2+3a^3\beta_3+a^4\beta_4\,.
\end{align}
Combining the above results, the entire $\fmn$ equation of motion 
becomes the following higher derivative equation for $\gmn$, 
\begin{align}\label{fEq2ndapp}
&\tfrac{\Omega}{a^2\alpha^2}\,\gmn+\tfrac{1}{\mu^2}
\left[1+\tfrac1{a^2\alpha^2}\right]\mathcal{G}_{\mu\nu}
+\tfrac{2\Omega}{a^2\alpha^2s_1\mu^2}\,P_{\mu\nu} 
+\tfrac{1}{\mu^4 s_1}B_{\mu\nu}\nn\\
&+\tfrac{\Omega}{a^2\alpha^2s_1^3\mu^4}\Big[(s_1+2s_2)P_\mu^{~\rho}
P_{\rho\nu} -2s_2PP_{\mu\nu}-\tfrac{s_2}{3}\gmn\left(P_{\rho\sigma}
P^{\rho\sigma}-P^2\right)\Big]
\nn\\
&-\tfrac{1}{s_1\mu^4}(1+\frac{1}{\alpha^2 a^2})\left[3P P_{\mu\nu}
-2P_\mu^{~\rho} P_{\rho\nu} -\frac{1}{2}g_{\mu\nu}(P^2-P^{\alpha\beta}
P_{\alpha\beta})\right]+\mathcal{O}\left(\tfrac{P^3}{\mu^6}\right) 
=0\,.
\end{align}
Here, some of the four-derivative terms have been collected
into the Bach tensor~\cite{Bach}, 
\be\label{bacht2}
B_{\mu\nu}=-\nabla^2P_{\mu\nu}&-\nabla_\mu\nabla_\nu P^\rho_{~\rho}
+\nabla_\rho\nabla_\mu P^\rho_{~\nu}+\nabla_\rho\nabla_\nu P^\rho_{~\mu}
-2P_\mu^{~\rho}P_{\rho\nu}+\tfrac1{2}\gmn P^{\rho\sigma}P_{\rho\sigma}\,,
\ee
which is invariant under local Weyl transformations of the metric.

\section{Four-derivative terms in gauge transformations}\label{appfd} 

Here we outline the computation of the four-derivative terms generated
by lower orders in the symmetry transformation. 

\subsection{Higher-curvature equations}\label{curvapp}

We start by presenting the explicit expressions for the
higher-curvature expansions of the equations of motion in the PM
bimetric model up to 10th order in derivatives. The equations read, 
\begin{subequations}
\label{symsola}
\begin{align}
\fmnp &= -\gmnp+4\rho\pmn+4\sum_{n=2}^\infty\rho^{n}\gamma_{\mu\nu}^{(2n)}[ g']\,,\label{symsolga}\\
\gmnp &= -\fmnp+4\rho\tilde{P}_{\mu\nu}+4\sum_{n=2}^\infty\rho^{n}\gamma_{\mu\nu}^{(2n)}[ f']\,, 
\qquad\quad
\rho\equiv\frac{\alpha}{4\beta_2\mu^2}\,.\label{symsolfa}
\end{align}
\end{subequations}
Here, the higher-derivative functions $\gamma_{\mu\nu}^{(2n)}$ for
$n\leq 5$ are of the form,\footnote{We have made use of the
  Cayley-Hamilton theorem~\eqref{CH} to reduce the tensor powers in
  $\gamma_{\mu\nu}^{(8)}$ and $\gamma_{\mu\nu}^{(10)}$.} 
\begin{align}
\gamma_{\mu\nu}^{(4)}[g]&=-2\pmn^2+e_1\pmn-\tfrac1{3}e_2\,\gmn\,,\\
\gamma_{\mu\nu}^{(6)}[g]&=4\pmn^3-5e_1\,\pmn^2+(e_1^2+2e_2)\pmn
-\left(e_3+\tfrac1{3}e_1e_2\right)\gmn\,,\\
\gamma_{\mu\nu}^{(8)}[g]&=8e_1\pmn^3-\left(9e_1^2-\tfrac{8}{3}e_2\right)\pmn^2
+\left(e_1^3+\tfrac{14}{3}e_1e_2-4e_3\right)\pmn\nn\\
&\hspace{170pt}\quad-\left(\tfrac{1}{3}e_1^2e_2-\tfrac{1}{9}e_2^2+\tfrac{8}{3}e_1e_3-\tfrac{20}{3}e_4\right)\gmn\,,\\
\gamma_{\mu\nu}^{(10)}[g]&=\left(13e_1^2-\tfrac{8}{3}e_2\right)\pmn^3-\left(14e_1^3-7e_1e_2-6e_3\right)\pmn^2\nn\\
&\hspace{120pt}\quad+\left(e_1^4+8e_1^2e_2+\tfrac{2}{3}e_2^2-13e_1e_3-8e_4\right)\pmn\nn\\
&\hspace{120pt}\quad-\left(\tfrac{1}{3}e_1^3e_2-\tfrac{1}{3}e_1e_2^2+5e_1^2e_3+\tfrac{1}{3}e_2e_3-22e_1e_4\right)\gmn\,.
\end{align}
In this expression all of the $e_n$ are elementary symmetric
polynomials of the Schouten tensor $P^{\mu}_{~\nu}=g^{\mu\rho}P_{\rho\nu}(g)$.

For future reference we also present the following linear variations
of the four-derivative terms, 
\begin{align}
\delta\gamma_{\mu\nu}^{(4)}=&\left[2P_\mu^{~\rho}P_\nu^{~\sigma}-\pmn P^{\rho\sigma}
-\frac1{3}e_2(P)\delta_\mu^\rho\delta_\nu^\sigma+\frac1{3}\gmn PP^{\rho\sigma}
-\frac1{3}\gmn P^\rho_{~\lambda}P^{\lambda\sigma}\right]\delta g_{\rho\sigma}\nn\\
&+\left[\pmn g^{\rho\sigma}+P\delta_\mu^\rho\delta_\nu^\sigma-2P_\mu^{~\rho}\delta_\nu^\sigma
-2P_\nu^{~\rho}\delta_\mu^\sigma-\frac1{3}\gmn P g^{\rho\sigma}+\frac1{3}\gmn P^{\rho\sigma}
\right]\delta P_{\rho\sigma}\,,
\end{align}
where the linearised Schouten tensor reads as,
\begin{align}\label{dP}
\delta P_{\mu\nu}&=\pmn[g+\delta g]-\pmn[g]\nn\\
&=-\frac1{2}\left[\nabla_\mu\nabla_\nu \delta g-\nabla^\rho\nabla_\mu \delta g_{\nu\rho}-\nabla^\rho\nabla_\nu \delta g_{\mu\rho}
+\nabla^2\delta g_{\mu\nu}\right]
+\frac1{6}\gmn\left[\nabla^2 \delta g-\nabla^\rho\nabla^\sigma \delta g_{\sigma\rho}\right]\nn\\
&\quad+\frac1{6}\gmn \delta g^{\rho\sigma}P_{\rho\sigma}
-\frac1{2}P^\rho_{~\rho}\left[\delta g_{\mu\nu}-\frac1{6}\gmn \delta g\right]\,,
\end{align}
and we have,
\be\label{dEnP}
\delta e_n(P) = \sum_{k=1}^n(-1)^ke_{n-k}(P)\left[[P^k]^{\rho\sigma}\delta g_{\rho\sigma}
-[P^{k-1}]^{\rho\sigma}\delta P_{\rho\sigma}\right]\,.
\ee
These will be useful for computing the contribution to the
four-derivative terms in the gauge transformations.

\subsection{Vanishing of the four-derivative contributions}

Consider arbitrary variations expanded in powers of derivatives up to
$4^{\mathrm{th}}$ order, 
\be
\Delta\gmnp&=&\Delta_{(0)}\gmnp+\Delta_{(2)}\gmnp+\Delta_{(4)}\gmnp+\hdots\,,\nn\\
\Delta\fmnp&=&\Delta_{(0)}\fmnp+\Delta_{(2)}\fmnp+\Delta_{(4)}\fmnp+\hdots\,.
\ee
The terms with zero and two derivatives were determined earlier,
\be
\Delta_{(0)}\gmnp&=&\phi\gmnp\,,\qquad
~~\,\Delta_{(2)}\gmnp=-2\rho\phi\pmn-2\rho\nabla_\mu\partial_\nu\phi\,,\nn\\
\Delta_{(0)}\fmnp&=&-\phi\gmnp\,,\qquad
\Delta_{(2)}\fmnp=2\rho\phi\pmn-2\rho\nabla_\mu\partial_\nu\phi\,.
\ee
Using~(\ref{symsolfa}) to replace $\gmnp=-\fmnp+4\rho\tilde{P}_{\mu\nu}+\hdots$, the transformations of $\fmnp$ can also be expressed in terms of $\fmnp$,
\be
\Delta_{(0)}\fmnp=\phi\fmnp\,,\qquad
\Delta_{(2)}\fmnp=-2\rho\phi\tilde{P}_{\mu\nu}-2\rho\tilde\nabla_\mu\partial_\nu\phi\,.
\ee
Demanding the invariance of~(\ref{symsolga}) at the four-derivative
level and using the results from appendix~\ref{curvapp}, we then
obtain the following relation among the transformations, 
\begin{align}\label{df_g}
\Delta_{(0)}\fmnp&+\Delta_{(2)}\fmnp+\Delta_{(4)}\fmnp\nn\\
&=-\phi\gmnp -\Delta_{(2)}\gmnp-4\rho\nabla_\mu\partial_\nu\phi
-\Delta_{(4)}\gmnp
-4\rho^2\gamma_{\mu\nu}^{(4)}[g']\phi\nn\\
&\quad-2\rho\Big[\nabla_\mu\nabla_\nu\Delta_{(2)}g'^\rho_{\,\rho}
-\nabla^\rho\nabla_\mu\delta \Delta_{(2)}g'_{\nu\rho}
-\nabla^\rho\nabla_\nu\delta \Delta_{(2)}g'_{\mu\rho}
+\nabla^2\Delta_{(2)}g'_{\mu\nu}\Big]\nn\\
&\quad+\tfrac{2}{3}\rho\,\gmnp\Big[\nabla^2\Delta_{(2)}g'^\rho_{\,\rho}
-\nabla^\rho\nabla^\sigma\Delta_{(2)}g'_{\rho\sigma}\Big]
+\tfrac{2}{3}\rho\,\gmnp P^{\rho\sigma}\Delta_{(2)}g'_{\rho\sigma}\nn\\
&\quad-2\rho P\Big[\Delta_{(2)}g'_{\mu\nu}
-\tfrac1{6}\gmnp\Delta_{(2)}g'^\rho_{\,\rho}\Big]
-4\rho^2\Big[\pmn\nabla^2\phi+P\nabla_\mu\nabla_\nu\phi\nn\\
&\quad-2P_\mu^{~\rho}\nabla_\rho\nabla_\nu\phi
-2P_\nu^{~\rho}\nabla_\rho\nabla_\mu\phi
-\tfrac1{3}\gmnp P\nabla^2\phi+\tfrac1{3}\gmnp P^{\rho\sigma}\nabla_\rho\nabla_\sigma\phi\Big]\,,
\end{align}
where all quantities on the right hand side are given in terms of
$\gmnp$. We now use~(\ref{symsolfa}) to express the right hand side of
\eqref{df_g} as a functional of $\fmnp$, 
\begin{align}\label{df_f}
\Delta_{(0)}\fmnp&+\Delta_{(2)}\fmnp+\Delta_{(4)}\fmnp\nn\\
&=~\phi\fmnp-4\rho\phi \tilde{P}_{\mu\nu}
-4\rho\tilde\nabla_\mu\partial_\nu\phi
+4\rho\, C_{\mu\nu}^{\ph{\mu\nu}\rho}\tilde\nabla_\rho\phi
-\left.\Delta_{(2)}\gmnp\right|_{g'=- f'+4\rho \tilde{P}}
+\left.\Delta_{(4)}\gmnp\right|_{g'=-f'}\nn\\
&\quad+2\rho\Big[\tilde\nabla_\mu\tilde\nabla_\nu\Delta_{(2)}g'^\rho_{\,\rho}
-\tilde\nabla^\rho\tilde\nabla_\mu\Delta_{(2)}g'_{\nu\rho}
-\tilde\nabla^\rho\tilde\nabla_\nu\Delta_{(2)}g'_{\mu\rho}
+\tilde\nabla^2\Delta_{(2)}g'_{\mu\nu}\Big]\nn\\
&\quad-\tfrac{2\rho}{3}\gmnp\Big[\tilde\nabla^2\Delta_{(2)}g'^\rho_{\,\rho}
-\tilde\nabla^\rho\tilde\nabla^\sigma\Delta_{(2)}\tilde{g}_{\rho\sigma}\Big]
-\tfrac{2\rho}{3}\gmnp \tilde{P}^{\rho\sigma}\Delta_{(2)}g'_{\rho\sigma}\nn\\
&\quad+2\rho \tilde{P}\Big[\Delta_{(2)}g'_{\mu\nu}
-\tfrac1{6}\gmnp\Delta_{(2)}g'^\rho_{\,\rho}\Big]
+4\rho^2\Big[\tilde{P}_{\mu\nu}\tilde\nabla^2\phi
+\tilde{P}\tilde\nabla_\mu\partial_\nu\phi\nn\\
&\quad
-2\tilde{P}_{\mu\rho}\tilde\nabla^\rho\partial_\nu\phi
-2\tilde{P}_{\nu\rho}\tilde\nabla^\rho\partial_\mu\phi
-\tfrac1{3}\gmnp \tilde{P}\tilde\nabla^2\phi+\tfrac1{3}\gmnp \tilde{P}^{\rho\sigma}\tilde\nabla_\rho\partial_\sigma\phi\Big]\,,
\end{align}
where now, on the right-hand side, all quantities (curvatures,
derivatives, index raising, etc.) are understood to be defined with
respect to $\fmnp$ and in the transformations $\gmnp$ is a function
of~$\fmnp$, 
\beqn
\Delta_{(2)}\gmnp=\left.\Delta_{(2)}\gmnp\right|_{g'=-f'+4\rho \tilde P}\,,\qquad
\Delta_{(4)}\gmnp=\left.\Delta_{(4)}\gmnp\right|_{g'=-f'}\,.
\eeqn
Furthermore, we have introduced
\be
C_{\mu\nu}^{\ph{\mu\nu}\rho}=-2\rho\left[\tilde\nabla_\mu \tilde{P}_\nu^{~\rho}+\tilde\nabla_\nu \tilde{P}_\mu^{~\rho}-\tilde\nabla^\rho \tilde{P}_{\mu\nu}\right]\,.
\ee
Next, we try to impose the interchange symmetry of the
transformations. According to our discussion in
section~\ref{symconstr}, this is only possible if the contributions to
the transformations that are generated from lower orders vanish. 
Indeed, if we demand,
\be
\Delta_{(4)}\fmnp=\left.\Delta_{(4)}\gmnp\right|_{g'\leftrightarrow f'}\,,
\ee
after a lengthy but straightforward calculation we find that this
condition becomes, 
\begin{align}\label{vancon}
\Delta_{(4)}\gmnp=\Delta_{(4)}\gmnp&+4\rho^2\Big[
\nabla_\rho\nabla_\mu\nabla_\nu\partial^\rho\phi+\nabla^\rho\nabla_\nu\nabla_\rho\partial_\mu\phi
-\nabla_\mu\nabla_\nu\nabla^2\phi-\nabla^2\nabla_\mu\partial_\nu\phi\Big]\nn\\
&-4\rho^2\Big[P\nabla_\mu\partial_\nu\phi+\nabla_\rho P_{\mu\nu}\partial^\rho\phi
+P_{\mu\rho}\nabla^\rho\partial_\nu\phi+P_{\nu\rho}\nabla^\rho\partial_\mu\phi\Big]\nn\\
&-\tfrac{4\rho}{3}\gmnp\Big[\nabla_\rho\nabla_\sigma\nabla^\rho\partial^\sigma\phi
-\nabla^2\nabla^2\phi-P^{\rho\sigma}\nabla_\rho\partial_\sigma\phi
-\tfrac1{2}P\nabla^2\phi\Big]\,.
\end{align}
Clearly, this leaves $\Delta_{(4)}\gmnp$ undetermined and instead
provides a consistency check on the possible existence of a symmetry
to this order in derivatives. Obviously, the condition is satisfied
for constant gauge parameter, but showing that the extra terms vanish
for a general function $\phi$ requires a bit more work. 

Some identities that are useful for our purposes at this point
are,\footnote{Note that our curvature conventions are
  $[\nabla_\mu,\nabla_\nu]\,\omega_\rho=R_{\mu\nu\rho}^{\ph{\mu\nu\rho}\sigma}\omega_\sigma$.} 
\begin{align}
\nabla_\rho\nabla_\mu\nabla_\nu\nabla^\rho\phi&=\nabla_\mu\nabla_\nu\nabla^2\phi
+R_{\rho\mu\nu}^{\ph{\rho\mu\nu}\sigma}\nabla_\sigma\nabla^\rho\phi
+P\nabla_\mu\nabla_\nu\phi+\frac1{2}\nabla_\mu P\nabla_\nu\phi\nn\\
&\qquad\qquad\qquad+P_\mu^{~\rho}\nabla_\nu\nabla_\rho\phi+P_\nu^{~\rho}\nabla_\mu\nabla_\rho\phi
+\nabla_\mu P_{\nu\rho}\nabla^\rho\phi\,,\\
\nabla^\rho\nabla_\nu\nabla_\rho\nabla_\mu\phi&=\nabla^2\nabla_\mu\nabla_\nu\phi
-R_{\rho\mu\nu}^{\ph{\rho\mu\nu}\sigma}\nabla_\sigma\nabla^\rho\phi
+\nabla^\rho R_{\nu\rho\mu}^{\ph{\nu\rho\mu}\sigma}\nabla_\sigma\phi\,,\\
\nabla_\rho\nabla_\sigma\nabla^\rho\nabla^\sigma\phi&=\nabla^2\nabla^2\phi+P^{\rho\sigma}\nabla_\rho\nabla_\sigma\phi
+\frac1{2}P\nabla^2\phi+\frac3{2}\nabla_\rho P\nabla^\rho\phi\,.
\end{align}
Using these identities, we can write the condition (\ref{vancon}) as,
\be\label{vancon2}
\nabla_\rho R_{\mu\sigma\nu}^{\ph{\mu\sigma\nu}\rho}\nabla^\sigma\phi
+\nabla_\mu P_{\nu\sigma}\nabla^\sigma\phi-\nabla_\sigma P_{\mu\nu}\nabla^\sigma\phi
+\frac1{2}\nabla_\mu P\nabla_\nu\phi-\frac1{2}\gmnp \nabla_\sigma P\nabla^\sigma\phi=0\,.
\ee
Now, in order to show that this is always true, consider the
(contracted) second Bianchi identity, 
\be
\nabla_\rho R_{\mu\sigma\nu}^{\ph{\mu\sigma\nu}\rho}+\nabla_\mu R_{\nu\sigma}-\nabla_\sigma R_{\mu\nu}=0\,,
\ee
or equivalently, using $R_{\mu\nu}=\pmn+\frac{1}{2}\gmn P$,
\be
\nabla_\rho R_{\mu\sigma\nu}^{\ph{\mu\sigma\nu}\rho}+\nabla_\mu P_{\nu\sigma}-\nabla_\sigma P_{\mu\nu}
+\frac1{2}g_{\nu\sigma}\nabla_\mu P-\frac1{2}g_{\mu\nu}\nabla_\sigma P=0\,.
\ee
Using this, it is obvious that (\ref{vancon2}) and hence
(\ref{vancon}) are identically satisfied for any function~$\phi$. This
proves that there are no four-derivative terms generated from lower
orders in the gauge transformations and hence the condition for the
existence of an invariance is satisfied at the first nontrivial
order.


\end{document}